\documentclass[shawpacs,aps,graphicx,twocolumn]{revtex4}
\usepackage{amsmath}
\usepackage{graphicx}
\begin{document}


\title{Quantum hyperentanglement and its applications in quantum information processing\footnote{Published
in Science Bulletin \textbf{62}(1), 46--68 (2017). }}
\author{Fu-Guo Deng$^{a,}$\footnote{email:
fgdeng@bnu.edu.cn}, Bao-Cang Ren$^{b,}$\footnote{email:
renbaocang@cnu.edu.cn}, and Xi-Han Li$^{c,d,}$\footnote{email:
xihanlicqu@gmail.com}}

\address{$^a$ Department of Physics, Applied Optics Beijing Area Major Laboratory,
Beijing Normal University, Beijing 100875, China\\
$^b$ Department of Physics, Capital Normal University, Beijing 100048, China\\
$^c$ Department of Physics, Chongqing University, Chongqing 400044, China\\
$^d$ Department of Physics and Computer Science, Wilfrid Laurier University,
Waterloo ON N2L 3C5, Canada}

\date{\today }

\begin{abstract}
Hyperentanglement is a promising resource in quantum information
processing with its high capacity character, defined as the
entanglement in multiple degrees of freedom (DOFs) of a quantum
system, such as polarization, spatial-mode, orbit-angular-momentum,
time-bin and frequency DOFs of photons. Recently, hyperentanglement
attracts much attention as all the multiple DOFs can be used to
carry information in quantum information processing fully. In this
review, we present an overview of the progress achieved so far in
the field of hyperentanglement in photon systems and some of its
important applications in quantum information processing, including
hyperentanglement generation, complete hyperentangled-Bell-state
analysis,  hyperentanglement concentration, and hyperentanglement
purification for high-capacity long-distance quantum communication.
Also, a scheme for hyper-controlled-not gate is introduced for
hyperparallel photonic quantum computation, which can perform two
controlled-not gate operations on both the polarization and
spatial-mode DOFs and depress the resources consumed and the
photonic dissipation.
\end{abstract}


\maketitle

\section{introduction}

Quantum information processing (QIP) has attracted considerable
interest and attention of scientists in a variety of disciplines
with its ability for improving the methods of dealing and
transmitting information \cite{NC,RMPadd}. Entanglement is a
distinctive feature of quantum physics \cite{HOR}, and it is very
useful in QIP, including both quantum communication and quantum
computation. Entangled photon systems are the natural resource for
establishing quantum channel in long-distance quantum communication,
especially in quantum repeaters \cite{repeater} for some important
tasks of communication, such as quantum key distribution
\cite{BB84,QKD2,QKD3}, quantum secret sharing \cite{QSS1}, and
quantum secure direct communication
\cite{QSDC1,QSDC2,QSDCDL04,DL04Exp,twostepExp}. In experiment, the
entangled photon systems are usually prepared by the spontaneous
parametric down-conversion (SPDC) process in nonlinear crystal
\cite{type2add,type11,type12}. In the conventional protocols for
quantum information processing, the entanglement in one degree of
freedom (DOF) of photon systems is selected in the SPDC process.  In
fact, there are more than one DOF in a quantum system, such as the
polarization, spatial-mode, orbit-angular-momentum, frequency, and
time-bin DOFs in  a  photon system.



Hyperentanglement, the simultaneous entanglement in multiple DOFs of
a quantum system, has been studied extensively in recent years. It
is a promising candidate for QIP with its high-capacity character.
In experiment, hyperentanglement can be generated by the combination
of the techniques used for creating entanglement in a single DOF
\cite{preparation1}. With this method, many different types of
hyperentangled states can be prepared
\cite{preparation2,preparation3,preparation4,preparation5,preparation6,preparation7,preparation9,preparation10},
such as the polarization-spatial hyperentangled state
\cite{preparation2}, polarization-spatial-time-energy hyperentangled
state \cite{preparation3}, and so on. Hyperentanglement is a
fascinating resource for quantum communication and quantum
computation. On one hand, it can assist us to implement  many
important tasks in quantum communication with one DOF of photons,
such as quantum dense coding with linear optics \cite{hcapcity}, the
complete Bell-state analysis for the quantum states in the
polarization DOF \cite{BSA,BSA1,BSA2,BSA3,EPPsheng2}, the
deterministic entanglement purification
\cite{EPPsheng2,EPPsheng3,EPPlixh,EPPdeng}, and the efficient
quantum repeater \cite{repeaterwangtj}. On the other hand,
hyperentanglement can be used directly in some important
applications in QIP. For example, it can improve the channel
capacity of quantum communication and speedup quantum computation
largely.

In the applications of hyperentanglement, the complete
hyperentangled-Bell-state analysis (HBSA)
\cite{HBSA,HBSA1,HBSA2,HBSA3,HBSA4,HBSA6,HBSALiuq,HBSALIXHOE,HBSA7,HBSAWW},
hyper-teleportation of quantum state with more than one DOF
\cite{HBSA},  hyperentanglement swapping \cite{HBSA1},
hyperentanglement concentration
\cite{HECP,HEPPECP,HECP1,HECP2,HECP4,HECP3,HECPLixhOE,HECPadd2},
hyperentanglement purification
\cite{HEPPECP,HEPP1,HEPP2wang,HEPPadd,HEPPaddWW}, and universal
entangling quantum gates for hyperparallel photonic quantum
computation
\cite{hypercnot,hypercnot1,h-hypercnot,hypercnot4,hypercnot5} are
very useful and important. HBSA is the prerequisite for
high-capacity quantum communication protocols with hyperentanglement
and it is used to distinguish the hyperentangled states.  Also, in
the practical application of hyperentanglement in quantum
communication, the hyperentangled photon systems are produced
locally, which leads to the decoherence of  the hyperentanglement
when the photons are distributed over a channel with environment
noise or stored in practical quantum devices. Quantum repeater is a
necessary technique to overcome the influence on quantum
communication from this decoherence \cite{repeater}. In
high-capacity quantum repeater with hyperentanglement,
hyperentanglement concentration and hyperentanglement purification
are two passive ways to recover the entanglement in nonlocal
hyperentangled photon systems. They are not only useful but also
absolutely necessary in long-distance high-capacity quantum
communication with hyperentanglement as the self-error-rejecting
qubit transmission scheme \cite{LIXHAPL} do not work in depressing
the influence of noise from both a long-distance channel and the
storage devices for quantum states. Moreover, quantum repeaters for
long-distance quantum communication require the entangled photons
with higher fidelity (usually $\sim$ 99\%) beyond that from faithful
qubit transmission schemes.

Different from conventional parallel quantum computation in which
the states of quantum systems in one DOF or equivalent are used to
encode information, hyperparallel photonic quantum computation
performs universal quantum gate operations on two-photon or
multi-photon systems by encoding all the quantum states of each
photon in multiple DOFs (two or more DOFs) as information carriers
\cite{hypercnot,hypercnot1,h-hypercnot,hypercnot4,hypercnot5}. With
hyperparallel photonic quantum logic gates, the resource consumption
can be reduced largely and the photonic dispassion noise can be
depressed in quantum circuit \cite{h-hypercnot}. Moreover, the
multiple-photon hyperentangled state can be prepared and measured
with less resource and less steps by using the hyperparallel
photonic quantum logic gates, which may speedup the quantum
algorithm \cite{hypercnot,hypercnot1}.

In this review, we will overview the development of
hyperentanglement and its applications in QIP in recent several
years. We will first review  the preparation of hyperentanglement,
and then introduce the applications of the hyperentanglement in
quantum communication, including hyper-teleportation of an unknown
quantum state  in more than two DOFs and hyperentanglement swapping.
We also highlight how to improve the entanglement of nonlocal
hyperentangled photon systems with hyperentanglement concentration
and hyperentanglement purification. At last, the principle of a
polarization-spatial hyper-controlled-not (hyper-CNOT) gate is described for hyperparallel
quantum computing.

\section{preparation of hyperentanglement}

Hyperentangled states offer significant advantages in QIP due to the
presence of quantum correlations in multiple DOFs. In this section,
we will introduce the preparation of hyperentangled states of photon
systems. In the first part, we overview the preparation of entangled
photon pairs with the SPDC process in nonlinear crystals. In the
second part, we overview the preparation of hyperentangled photon
systems with the combination of the techniques used for creating
entanglement in single DOF.

\subsection{Preparation of entanglement in single DOF}

Generally speaking, the most extensive method used to generate an
entangled state is the SPDC
process in a nonlinear crystal. When a pump laser beam $p$ shines a
nonlinear birefringent crystal, the idler photon $i$ and the signal
photon $s$ are generated probabilistically from the crystal. The
maximal probability can be achieved by satisfying two matching
conditions. One is the phase-matching:
\begin{eqnarray}  
\vec{k}_p=\vec{k}_s+\vec{k}_i,
\end{eqnarray}
and the other is energy-matching:
\begin{eqnarray}  
\omega_p=\omega_s+\omega_i.
\end{eqnarray}
Here $\vec{k}$ represents the wave vector and $\omega$ denotes the
frequency. Usually, there are two common kinds of phase-matching
adopted in experiment, depending on the extraordinary $(e)$ and the
ordinary $(o)$ polarizations of the pump photon and the two SPDC
photons. The type-I phase-matching is $e \rightarrow o+o$ and the
type-II phase-matching is $e \rightarrow e+o$.

\begin{center}
\begin{figure}[!h]
\includegraphics*[width=7.2cm]{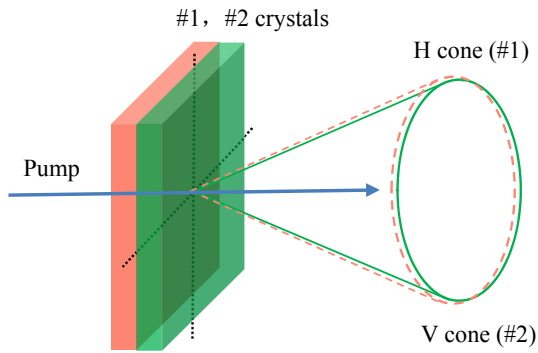}   
\caption{Type-I polarization entanglement sources
\cite{type11}.}\label{type1}
\end{figure}
\end{center}

In the type-I phase-matching, two SPDC photons are both ordinary and
have the same polarizations. To generate an entangled state, two
crystals with orthogonal optical axes can be used \cite{type11}. The
principle is shown in Fig.~\ref{type1}. To satisfy the
phase-matching condition, two correlated photons are emitted over
opposite directions of the cone surface. By selecting one pair of
the correlated wavevector modes, the polarization entangled states
$\vert \Phi^{\pm}\rangle=\frac{1}{\sqrt{2}}(\vert HH\rangle \pm\vert
VV\rangle)$ can be prepared. Here $H$ and $V$ represent the
horizontal and vertical polarization states of a photon,
respectively. An alternative way to prepare an entangled state with
type-I phase-matching is using a single crystal and a double passage
of the laser beam after reflection on a mirror \cite{type12}.

\begin{center}
\begin{figure}[!h]
\includegraphics*[width=8cm]{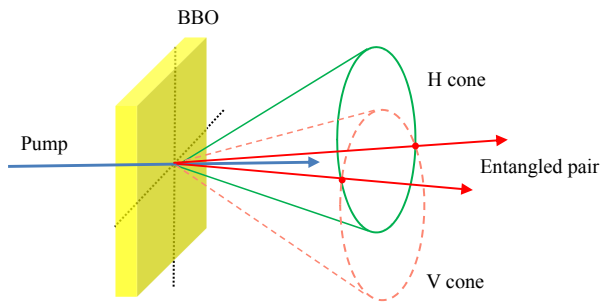}   
\caption{Non-collinear type-II phase-matching spontaneous parametric
down-conversion process \cite{type2add}.}\label{type2}
\end{figure}
\end{center}

In the type-II phase-matching, the two degenerate photons are
emitted over two different mutually crossing emission cones
\cite{type2add}. The emission directions of the signal and idler
photons are symmetrically oriented with respect to the propagation
direction of the pump photon. The two entangled photons are
generated along the direction of the intersection of the two cones.
Since the ordinary and extraordinary photons have orthogonal
polarization states, the polarization entangled states $\vert
\Psi^{\pm}\rangle=\frac{1}{\sqrt{2}}(\vert HV\rangle \pm\vert
VH\rangle)$ are prepared with type-II phase-matching. If the two
cones only intersect at one point, it is called the collinear SPDC
process and orthogonally polarized photons are indistinguishable at
exactly this point. The type-II collinear down-conversion is more
commonly used in experiment, as it offers a trivial way to
deterministically separate the photon pair by their polarization and
to work with each photon separately. For the non-collinear type-II
SPDC process which is shown is Fig.~\ref{type2}, the two emission
cones have two intersection directions, which can be made
indistinguishable with respect to their polarization, and then the
entangled state is generated.

Actually, by obeying these two matching conditions, entanglement in other
DOFs can be prepared, such as frequency, time-bin, and spatial-mode DOFs. The
SPDC photon pairs are coherently emitted at different emission times
as long as the interaction time of the pump wave with the crystal is
shorter than the coherence time of the pump photons. The photons are
automatically generated into an energy-entangled state due to the
nature of the SPDC process. In a word, the energy-time correlations are
presented in all SPDC photon pairs. The spatial-mode entangled state can be
generated by selecting more correlated directions. And if the
frequency of the idler and signal photons are not the same, they are
always entangled to fulfill the energy-matching condition.

Besides spin, photons possess a further angular momentum, the
orbital-angular-momentum (OAM), described by the Laguerre-Gaussian
mode $\vert l,p\rangle$. Under the collinear phase-matching
conditions, the OAM of these photons should satisfy $l_p=l_s+l_i$
(consider simple situation $p_i=p_s=0$). Therefore, when the pump
beam is a Gaussian TEM$_{00}$ beam, the two generated photons have
opposite $l$ as
\begin{eqnarray}  
\vert \Psi\rangle=\sum_{l=-\infty}^{+\infty}\sqrt{P_l}\,\vert
l\rangle_s\vert -l\rangle_i,
\end{eqnarray}
which is an OAM entangled state. Here $\sqrt{P_l}$ denotes the
probability of creating a signal photon with OAM $l$ and an idler
one with $-l$.

\subsection{Hyperentanglement in more than one degree of freedom}

\begin{center}
\begin{figure}[!h]
\includegraphics*[width=1.5in]{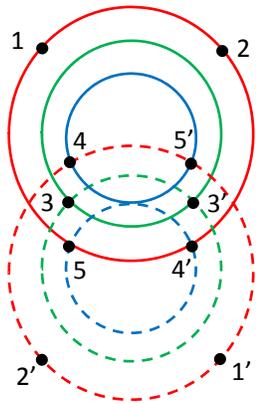}    
\caption{Schematic diagram of the type-II phase-matching curves. The
upper half (solid lines) are three extraordinary-polarized cones
with different wavelengths  \cite{preparation1}. The
ordinary-polarized cones are shown in the lower half (dashed lines)
with corresponding wavelength. The numbered points represent the
areas of directions to extract hyperentangled states.}\label{hyper1}
\end{figure}
\end{center}

The techniques used for creating single DOF entanglement can be
combined to generate hyperentanglement, which is entangled in more
than one DOF in the same time. The first proposal of an
energy-momentum-polarization hyperentangled state with a type-II
phase-matching was presented by Kwiat \cite{preparation1} in 1997.
The schematic diagram is shown in Fig.~\ref{hyper1}. The photons
emitted from conjugate points are all energy-time entangled. And the
photons generated from $3$ and $3'$ are automatically in a
polarization entangled state since these two cones have opposite
polarizations, which are indistinguishable at $3$ and $3'$. However,
these photons have definite momentum. Photons emitted from
$1-1'-2-2'$ are entangled in momentum with the definite polarization
state. The quantum state with photons generated along the directions
$4-4'-5-5'$ are entangled in momentum, energy-time, and polarization
simultaneously.

\begin{center}
\begin{figure}[!h]
\includegraphics*[width=3.0cm]{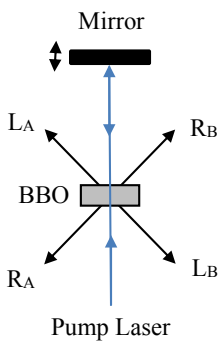}   
\caption{Schematic diagram of the setup to generate
hyperentanglement in both polarization and spatial-mode DOFs
\cite{preparation2}.}\label{hyper2}
\end{figure}
\end{center}

In 2005, Yang et al. \cite{preparation2} generated a two-photon state entangled
both in polarization and spatial-mode DOFs to realize the
all-versus-nothing test of local realism. The setup of generation is
shown in Fig.~\ref{hyper2}. In their experiment, the pump pulse
passes through the nonlinear BBO ($\beta$-barium borate) crystal
twice. The first passage of laser prepares a polarization-entangled
pairs $\vert \Psi^-\rangle_p=\frac{1}{\sqrt{2}}(\vert
H\rangle_A\vert V\rangle_B-\vert V\rangle_A\vert H\rangle_B)$ in the
spatial modes $L_A$ and $R_B$ with a small probability.  Then the
pump beam is reflected by a mirror and goes through the crystal a
second time (again). In this time, it probabilistically generates a $\vert
\Psi^-\rangle_p$ state in another two path modes $R_A$ and $L_B$.
The generation probabilities of two passages can be adjusted to equal.
Therefore, if there is perfect temporal overlap of modes $R_A$ and
$L_A$ ($R_B$ and $L_B$), the two possible ways of producing may
interfere, which results a spatial mode entangled state $\vert
\Psi^-(\phi)\rangle_s=\frac{1}{\sqrt{2}}(\vert R\rangle_A\vert
L\rangle_B-{\rm e}^{{\rm i}\phi}\vert L\rangle_A\vert R\rangle_B)$. $\phi=0$ can
be achieved by adjusting the distance between the mirror and the
crystal. Then, a maximally hyperentangled state in both polarization and
spatial mode ($\vert \Psi^-\rangle_p\otimes\vert\Psi^-(0)\rangle_s$)
is generated. In their experiment, the generation rate of entangled
photon pairs achieves $3.2\times 10^4$ per second.

In the same year, an experimental demonstration of a photonic
hyperentangled system which simultaneously entangled in
polarization, spatial-mode, and time-energy was reported
\cite{preparation3}. In their experiment, the entangled pairs are
prepared with the type-I phase-matching, and two BBO crystals with
orthogonal optical axes are used, which produce pairs of
horizontally and vertically polarized photons, respectively. Since
the spatial modes emitted from each crystal are indistinguishable,
the photon pairs are polarization entangled. Moreover, photon pairs
from a single nonlinear crystal are entangled in OAM. And according
to the energy-matching condition, each pair is entangled in energy
too. The generated state can be written as
\begin{eqnarray}  
&&(\vert HH\rangle+\vert VV\rangle) \otimes (\vert
+1,-1\rangle+\alpha \vert 0,0\rangle+ \vert -1,+1\rangle)\nonumber\\
&&\otimes (\vert SS\rangle+\vert LL\rangle).
\end{eqnarray}
Here $\vert \pm1\rangle$ and $\vert 0\rangle$ represent the
Laguerre-Gauss modes carrying $\pm\hbar$ and  0 OAM, respectively.
$\vert S\rangle$ and $\vert L\rangle$ denote the relative early and
late emission time of photons, respectively. $\alpha$ indicates the
OAM spatial-mode balance prescribed by the source and selected via
the mode-matching conditions. By collecting only $\pm \hbar$ OAM
state, the state in the spatial subspace is also a Bell state. The
total dimension of this hyperentangled system is
$2\times2\times3\times3\times2\times2=144$. To verify the quantum
correlations, they tested each DOF against the
Clauser-Horne-Shimony-Holt (CHSH) Bell inequality, and the results
showed for each DOF the Bell parameter exceeded the classical limit.
They also fully characterized the polarization and spatial-mode
state $(2\otimes2\otimes3\otimes3)$ subspace by tomography and
obtained the maximum fidelity $F=0.974$.

\begin{center}
\begin{figure}[!h]
\includegraphics*[width=6cm]{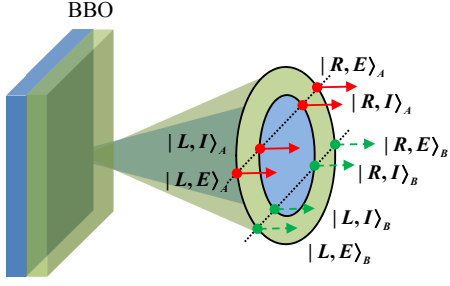}   
\caption{Schematic diagram of generating hyperentanglement in both
polarization and two longitudinal-momentum DOFs \cite{preparation4}.
Two BBO crystal slabs are used to generate hyperentangled photon
pairs $A$ (solid line) and $B$ (dashed line). $E$ and $I$ denote the
external and internal cones, respectively. $R$ and $L$ refer to the
left and right sides of each cone, respectively. For simplicity, the
positive lens which transforms the conical parametric emission of
the crystal into a cylindrical one is not shown in the figure.
}\label{hyper4}
\end{figure}
\end{center}

In 2009, Vallone et al. \cite{preparation4} also realized a
two-photon six-qubit hyperentangled state which is entangled in
polarization and two longitudinal momentum DOFs. The system used to
generate the state consists of two type-I BBO crystal slabs. The
polarization entanglement is created by spatially superposing the
two perpendicularly polarized emission cones of each crystal. Since
the two nonlinear crystal are cut at different phase matching
angles, the photon pairs will be created along the surfaces of two
cones, called the ``internal" ($I$) and ``external" ($E$) ones.
Coherence and indistinguishability between these two emission cones
are guaranteed by the coherence length of the pump beam. The double
longitudinal momentum entanglement is generated by singling out four
pairs of correlated modes with an eight-hole screen, shown in
Fig.~\ref{hyper4}. The hyperentangled state, which is a product of
one polarization entanglement and two longitudinal momentum
entanglement, can be written as
\begin{eqnarray}  
\frac{1}{\sqrt{2}}\left(\vert H\rangle_A\vert
H\rangle_B+{\rm e}^{{\rm i}\phi_1}\vert V\rangle_A\vert V\rangle_B\right)\nonumber\\
\otimes \frac{1}{\sqrt{2}}\left(\vert L\rangle_A\vert
R\rangle_B+{\rm e}^{{\rm i}\phi_2}\vert R\rangle_A\vert L\rangle_B\right)\nonumber\\
\otimes \frac{1}{\sqrt{2}}\left(\vert I\rangle_A\vert
I\rangle_B+{\rm e}^{{\rm i}\phi_3}\vert E\rangle_A\vert E\rangle_B\right).
\end{eqnarray}
Here $L$ and $R$ refer to the left and right sides of each cone. The
three relative phases $\phi_1$, $\phi_2$, and $\phi_3$ can be adjusted
in experiment.

\begin{center}
\begin{figure}[!h]
\includegraphics*[width=6cm]{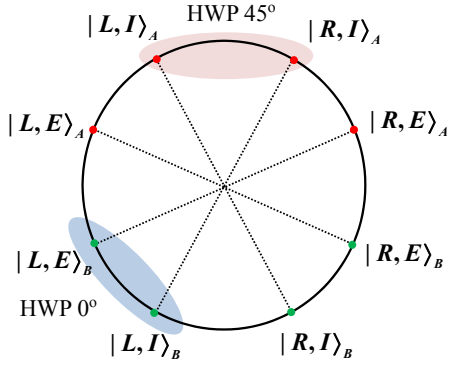}   
\caption{Model labeling of the hyperentangled state prepared with
type-I BBO crystal \cite{preparation5}. To get the six-qubit cluster
state, HWPs oriented at $45^o$ are placed on the $I$ modes of $A$
and HWPs oriented at $0^o$ are placed on the $L$ modes of $B$.
}\label{hyper5}
\end{figure}
\end{center}

The hyperentangled states prepared in the previous protocols are the
product states of different entangled DOFs. In 2009, Ceccarelli et
al. \cite{preparation5}  generated a two-photon six-qubit linear
cluster state by transforming a two-photon hyperentangled state
which is originally entangled in polarization and two linear
momentum DOFs. First, they generated the following six-qubit
hyperentangled state $\vert \widetilde{HE}_6\rangle$ by SPDC in a
Type-I BBO crystal, shown in Fig.~\ref{hyper5}.
\begin{eqnarray}  
\vert \widetilde{HE}_6\rangle &=&\frac{1}{\sqrt{2}}\left(\vert
E\rangle_{A}\vert E\rangle_{B}+\vert I\rangle_{A}\vert I\rangle_{B}\right)\nonumber\\
&& \otimes \frac{1}{\sqrt{2}}\left(\vert H\rangle_{A} \vert
H\rangle_{B}-\vert V\rangle_{A}\vert V\rangle_{B}\right)\nonumber\\
&& \otimes \frac{1}{\sqrt{2}}\left(\vert L\rangle_{A}\vert
R\rangle_{B}+\vert R\rangle_{A}\vert L\rangle_{B}\right).
\end{eqnarray}
Here $A$ and $B$ correspond to the up and down sides of the emission
cones, respectively. By encoding the qubits 1 and 4 with $E/I$ DOF,
qubits 2 and 5 with $H/V$ DOF, and qubits 3 and 6 with $R/L$ DOF,
the desired two-photon six-qubit linear cluster state $\vert
\widetilde{LC}_6\rangle$ is described as
\begin{eqnarray}  
\vert \widetilde{LC}_6\rangle&=&\frac{1}{2}\big[\vert
EE\rangle_{AB}\vert \Phi^-\rangle\vert LR\rangle_{AB}+\vert
EE\rangle_{AB}\vert \Phi^+\rangle\vert RL\rangle_{AB}\nonumber\\ &&
-\vert II\rangle_{AB}\vert \Psi^-\rangle\vert LR\rangle_{AB}+\vert
II\rangle_{AB}\vert \Psi^+\rangle\vert RL\rangle_{AB}\big].\nonumber\\
\end{eqnarray}
The transformation from the hyperentangled state to the cluster
state is carried out by  applying half-wave plates (HWPs) oriented
at $45^o$ on the internal $A$ modes and HWPs oriented at $0^o$ on
the left $B$ modes. The fidelity of the generated state is measured
and $F=0.6350\pm0.0008$ is obtained, which is 7 \% better than the
best previous result for six-qubit graph state with six particles.
The characterization and application of this state \cite{preparation6} were also
investigated in 2010.

\begin{center}
\begin{figure}[!h]
\includegraphics*[width=8cm]{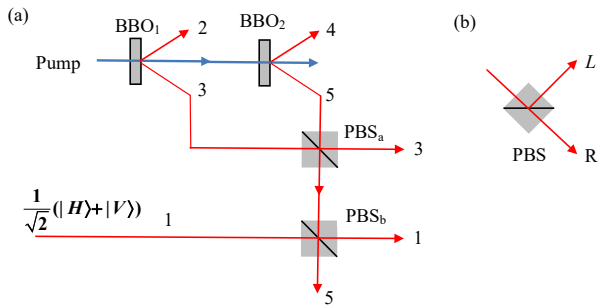}    
\caption{Experiment setup for generating ten-qubit five-photon
hyperentangled state \cite{preparation7}. (a) Generation of five-photon polarization
entangled state with postselection. (b) Creation of polarization-spatial
mode entanglement with PBS. $L$ and $R$ are two different spatial
modes.}\label{hyper7}
\end{figure}
\end{center}

Later, the hyperentangled state has been extended to ten-qubit
Schr\"odinger cat state in experiment, which carries the genuine
multi-qubit entanglement \cite{preparation7}. Although the previous
schemes demonstrated the hyperentangled states with really high
dimensions, they were only focused on two-photon states. In 2010,
Gao et al.  \cite{preparation7}  generated the genuine multipartite
hyperentanglement. The generation in their demonstration is composed
of two steps. In the first step, a five-photon polarization
entangled cat state is prepared by post-selection. Two pairs of
entangled photons are produced by SPDC in the state $(\vert
HH\rangle+\vert VV\rangle)/\sqrt{2}$ and a single photon  is
prepared in $(\vert H\rangle+\vert V\rangle)/\sqrt{2}$ state. The
principle of the first step is shown in Fig.~\ref{hyper7}. Two
polarizing beam splitters (PBSs), which transmit the horizontal
state $\vert H\rangle$ and reflect the vertical state $\vert
V\rangle$, are used to post-select the five-photon cat state $\vert
Cat\rangle^5_p=(\vert H\rangle^{\otimes 5}+\vert V\rangle^{\otimes
5})/\sqrt{2}$. In detail, the situation that each numbered spatial
mode has one and only one photon kept, which corresponds to the
desired state. Then, each photon is guided to a PBS, and the
ten-qubit hyperentangled state is produced:
\begin{eqnarray}  
\vert Cat \rangle^{10}=\frac{1}{\sqrt{2}}\left(\vert
H\rangle^{\otimes 5}\vert R\rangle^{\otimes 5}+\vert
V\rangle^{\otimes 5}\vert L\rangle^{\otimes 5}\right).
\end{eqnarray}
Here $L$ and $R$ signify the two spatial modes of each photon.

So far, the generation of entangled state is implemented via the nonlinear
optical process of the SPDC in different types of nonlinear crystals.
Recently, some works have been focused on waveguides due to their
high efficiency and on-chip integratability. In 2014, a
hyperentangled photon source in semiconductor waveguides was
proposed and demonstrated, which offers an alternative  path to realize an
electrically pumped hyperentangled photon source
\cite{preparation9}. They utilized phase-matching in Bragg
reflection waveguides to produce hyperentangled pairs through two
type-II SPDC processes. The ideal hyperentangled state in mode and
polarization DOFs is
\begin{eqnarray}  
\frac{1}{2}\left(\vert H,V\rangle+{\rm e}^{{\rm i}\phi} \vert V,
H\rangle)\otimes (\vert B,T\rangle+{\rm e}^{{\rm i}\psi}\vert T,B\rangle\right).
\end{eqnarray}
Here $T$ and $B$ denote the total internal reflection (TIR) mode and
Bragg mode, respectively. The fully entangled fraction of the
generated state is calculated, whose maximum value can achieve 0.99.

\begin{center}
\begin{figure}[!h]
\includegraphics*[width=8cm]{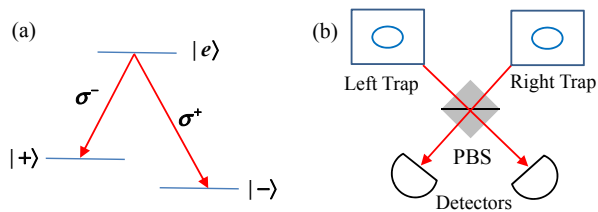}   
\caption{Experiment principle for generating hyperentangled states
between  atomic ions  \cite{preparation10}. (a) Internal energy
levels of ion. The $\vert +\rangle$ and $\vert -\rangle$ are ground
states and $\vert e\rangle$ is the excited state. $\sigma^+$ and
$\sigma^-$ are the polarized photons emitted due to the energy
decay. (b) The setup of generating spin entanglement between two
atomic ions confined in separated Paul traps.}\label{hyper10}
\end{figure}
\end{center}

Usually, the entanglement can only be generated locally. Since the
photon is one of the most ideal candidates for quantum
communication, most of the previous hyperentanglement generation
schemes are based on photons. Actually, other physical entity can
also be used as the carrier of hyperentanglement. For example, Hu et
al. \cite{preparation10} proposed a scheme of generating four-qubit
hyperentangled state between a pair of distant non-interacting
atomic ions which are confined in Paul traps. The state is entangled
in both spin and motion DOFs. The atomic ions with a $\Lambda$
configuration move along one direction in the Paul trap. The
principle for generating hyperentangled states between atomic ions
is shown in Fig.~\ref{hyper10}. First, the ions in each trap are
excited to the excited state $\vert e\rangle $ in spin DOF. Then
they decay along two possible channels $\vert e\rangle\rightarrow
\vert +\rangle$ and $\vert e\rangle\rightarrow \vert -\rangle$
accompanied by the emission of a $\sigma^-$ (with the spin $-1$) or
$\sigma^+$  (with the spin $+1$) polarized photon, respectively.
Therefore, the system consisting of the spin state of the ion and
the polarization of the emitting photon is in a maximally entangled
state:
\begin{eqnarray}  
\vert \Psi_1\rangle=\frac{1}{\sqrt{2}}\left(\vert +\rangle\vert
\sigma^-\rangle+\vert -\rangle\vert \sigma^+\rangle\right).
\end{eqnarray}
Then the two emitting photons from two traps are guided to a PBS. By
post-selecting the case that each atom emits a single photon, the two ions are
entangled in spin DOF as
\begin{eqnarray}  
\vert \Psi_2\rangle=\frac{1}{\sqrt{2}}\left(\vert +\rangle_{\rm L}\vert
-\rangle_{\rm R}+\vert -\rangle_{\rm L}\vert +\rangle_{\rm R}\right).
\end{eqnarray}
Here $L$ and $R$ denote the left and the right traps, respectively, as shown in
Fig.~\ref{hyper10}b. The motion DOF of both ions is initially in the ground
state $\vert 0\rangle_{\rm L}\vert 0\rangle_{\rm R}$. Then the entanglement is
transferred to the motion DOF with a sequence of laser pulses.
\begin{eqnarray}  
\vert \Psi_3\rangle=\frac{1}{\sqrt{2}}\vert +\rangle_{\rm L}\vert
+\rangle_{\rm R}(\vert 0\rangle_{\rm L}\vert 0\rangle_{\rm R}-\vert 1\rangle_{\rm L}\vert
1\rangle_{\rm R}).
\end{eqnarray}
Finally, by repeating the first step, the following hyperentangled
state can be produced,
\begin{eqnarray}  
\vert \Phi\rangle=\frac{1}{2}(\vert +\rangle_{\rm L}\vert -\rangle_{\rm R}+\vert
-\rangle_{\rm L}\vert +\rangle_{\rm R})\otimes(\vert 0\rangle_{\rm L}\vert
0\rangle_{\rm R}-\vert 1\rangle_{\rm L}\vert 1\rangle_{\rm R}).\nonumber\\
\end{eqnarray}

This proposal is experimentally feasible, although it has not been
demonstrated in  labs.

\section{High-capacity quantum communication with hyperentanglement}

\subsection{Status of Bell-state analysis for photonic quantum systems}

Bell-state analysis (BSA), which is used to distinguish the four
orthogonal Bell states of a two-particle quantum system  in one DOF,
is the prerequisite for quantum communication protocols with
entanglement and it is one of the important parts in quantum
repeaters. In 1999, two linear optical BSA protocols were proposed
by Vaidman's \cite{l-BSA} and L\"{u}tkenhau's \cite{l-BSA1} groups,
respectively, where the success probability is 50\%.  When
hyperentanglement is used to assist the analysis of Bell states, one
can completely distinguish all the four Bell states of a two-photon
system in one DOF. For example, in 1998, Kwiat and Weinfurter
\cite{BSA} proposed two complete BSA protocols by using the
hyperentanglement, which can distinguish the four orthogonal Bell
states in polarization DOF with the success probability 100\%. In
2003, Walborn et al.  \cite{BSA1} presented two complete BSA
protocols for photon pairs entangled in one DOF with
hyperentanglement, resorting to linear optical elements. Schuck et
al.  \cite{BSA2} and Barbieri  et al. \cite{BSA3} demonstrated the
complete BSA protocols in experiment by assisting hyperentanglement.


In high-capacity long-distance quantum communication, HBSA is also
required to attach some important goals, especially in high-capacity
quantum repeaters, teleportation of an unknown quantum state in two
or more DOFs, and hyperentanglement swapping. In 2007, Wei et al.
\cite{HBSA2} proposed a HBSA protocol with linear optical elements,
which can only distinguish 7 hyperentangled Bell states from 16
hyperentangled Bell states. In order to completely distinguish the
16 hyperentangled Bell states, nonlinear optical elements are
required.

The complete HBSA originates from the work by Sheng  et al.
\cite{HBSA} in 2010. They proposed the first scheme for the complete
HBSA of the two-photon polarization-spatial hyperentangled states
with cross-Kerr nonlinearity and designed the pioneering model for
teleporting an unknown quantum state in more than one DOF. In 2012,
Ren et al. \cite{HBSA1} introduced another interesting scheme for
the complete HBSA for photon systems by using the giant nonlinear
optics in quantum-dot-cavity systems and presented the
hyperentanglement swapping with photonic polarization-spatial
hyperentanglement. In 2012, Wang et al. \cite{HBSA3} presented an
important scheme for the complete HBSA for photon systems by the
giant circular birefringence induced by double-sided
quantum-dot-cavity systems. In 2015, Liu and Zhang \cite{HBSA4}
proposed two important schemes for hyperentangled-Bell-state
generation and HBSA assisted by nitrogen-vacancy (NV) centers in
resonators. Li and Ghose \cite{HBSA6} presented a very simple scheme
for the self-assisted complete maximally hyperentangled state
analysis via the cross-Kerr nonlinearity and another interesting
HBSA scheme \cite{HBSALIXHOE} for polarization and time-bin
hyperentanglement. Up to now, there are several important schemes
for the analysis of hyperentangled states
\cite{HBSA,HBSA1,HBSA2,HBSA3,HBSA4,HBSA6,HBSALiuq,HBSALIXHOE,HBSA7,HBSAWW},
including the probabilistic one based on linear optical elements
\cite{HBSA2} and the one for hyperentangled
Greenberger-Horne-Zeilinger (GHZ) states \cite{HBSA7}. In 2015, Wang
et al. \cite{hypereleportExp} demonstrated in experiment the quantum
teleportation of an unknown quantum state of a single photon in
multiple DOFs  by implementing the HBSA of two-photon systems
probabilistically with linear optical elements and ancillary
entanglement sources.  In 2016, Liu et al. \cite{HBSALiuq} gave the
original scheme for the complete nondestructive analysis of
two-photon six-qubit hyperentangled Bell states assisted by
cross-Kerr nonlinearity.

Here, we will introduce two high-capacity quantum communication
protocols, including teleportation of an unknown quantum state of a single
photon in two DOFs with hyperentanglement \cite{HBSA} and
hyperentanglement swapping \cite{HBSA,HBSA1}. First, we introduce
the complete HBSA for the polarization and spatial-mode DOFs of
photon systems \cite{HBSA}, which is an important technique in
high-capacity long-distance quantum communication. In the second
part, we introduce the quantum teleportation protocol based on
polarization-spatial hyperentanglement  \cite{HBSA}. At last, a
hyperentanglement swapping protocol  \cite{HBSA,HBSA1} is introduced
for quantum repeater and quantum communication.

\subsection{Hyperentangled Bell-state analysis}

The polarization-spatial hyperentangled Bell state is defined as the
two-photon system entangled in both the polarization and
spatial-mode DOFs, such as $|\Phi^+\rangle^{AB}_{PS}
=\frac{1}{2}(|HH\rangle+|VV\rangle)^{AB}_P\otimes
(|a_1b_1\rangle+|a_2b_2\rangle)^{AB}_S$. Here the superscripts $A$
and $B$ represent the two photons, and the subscripts $P$ and $S$
represent the polarization and spatial-mode DOFs, respectively.
$|i_1\rangle$ and $|i_2\rangle$ represent the two spatial modes of
photon $i$ ($i=A,B$). The polarization Bell states and the
spatial-mode Bell states are defined as
\begin{equation}   
\begin{split}
|\phi^\pm\rangle^{AB}_{P} &=
\frac{1}{\sqrt{2}}(|HH\rangle\pm|VV\rangle)^{AB}_{P}, \\
|\psi^\pm\rangle^{AB}_{P} &=
\frac{1}{\sqrt{2}}(|HV\rangle\pm|VH\rangle)^{AB}_{P}, \\
|\phi^\pm\rangle^{AB}_{S} &=  \frac{1}{\sqrt{2}}(|a_1b_1\rangle\pm|a_2b_2\rangle)^{AB}_{S}, \\
|\psi^\pm\rangle^{AB}_{S} &=
\frac{1}{\sqrt{2}}(|a_1b_2\rangle\pm|a_2b_1\rangle)^{AB}_{S},
 \end{split}
\end{equation}
where $|\psi^\pm\rangle^{AB}_{P}$ and $|\psi^\pm\rangle^{AB}_{S}$ are the
Bell states in the odd-parity mode, and $|\phi^\pm\rangle^{AB}_{P}$ and
$|\phi^\pm\rangle^{AB}_{S}$ are the  Bell states in the even-parity mode.
The 16 orthogonal hyperentangled Bell states can be distinguished completely by using
polarization parity-check quantum nondemolition detectors (QNDs) and spatial-mode
parity-check QNDs, assisted by cross-Kerr nonlinearity.

\begin{center}
\begin{figure}[!h]
\includegraphics*[width=8cm]{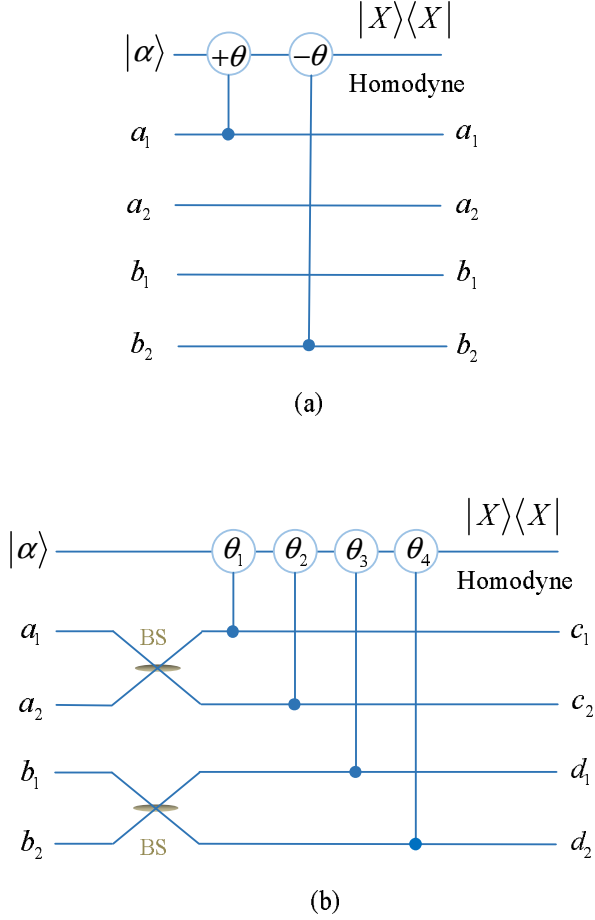}   
\caption{Schematic diagram of the HBSA for the spatial-mode Bell
states, resorting to the cross-Kerr nonlinearities \cite{HBSA}. (a)
The spatial-mode parity-check QND for distinguishing the states
$|\phi^\pm\rangle^{AB}_{S}$ from $|\psi^\pm\rangle^{AB}_{S}$. (b)
The second step for the analysis of spatial-mode Bell states, which
is used to distinguish the state $|\psi^+\rangle^{AB}_S$
($|\phi^+\rangle^{AB}_S$) with the relative phase shift $0$ from the
state $|\psi^-\rangle^{AB}_S$ ($|\phi^-\rangle^{AB}_S$) with the
relative phase shift $\pi$. BS represents a 50:50 beam splitter,
which is used to perform a Hadamard operation on the spatial-mode
DOF of a photon.}\label{HBSA1}
\end{figure}
\end{center}

\begin{center}
\begin{figure}[!h]
\includegraphics*[width=8cm]{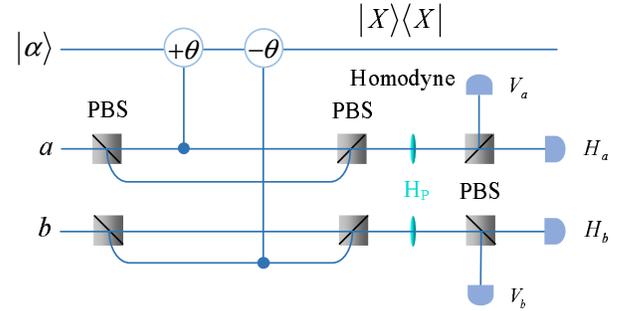}   
\caption{Schematic diagram of the HBSA for the polarization Bell
states, resorting to the cross-Kerr nonlinearities \cite{HBSA}. PBS
represents a polarizing beam splitter, which transmits the photon in
the horizontal polarization $\vert H\rangle$ and reflects the photon
in the vertical polarization $\vert V\rangle$, respectively. H$_P$
represents a half-wave plate which is used to perform a Hadamard
operation on the polarization DOF of a photon. $H_i$ or $V_i$
($i=a,b$) represents a single-photon detector.}\label{HBSA2}
\end{figure}
\end{center}

The Hamiltonian of cross-Kerr nonlinearity is described as
$H_{\rm ck}=\hbar\chi a^\dag_sa_sa^\dag_pa_p$, where $\hbar\chi$
represents the coupling strength of the nonlinear material.
$a^\dag_s$ ($a_s$) and $a^\dag_p$ ($a_p$) are the creation
(annihilation) operators. With this cross-Kerr interaction, the
system composed of a single photon and a coherent state can be
evolved as
\begin{equation}  
(a|0\rangle+b|1\rangle)|\alpha\rangle\;\;\rightarrow\;\;
a|0\rangle|\alpha\rangle+b|1\rangle |\alpha {\rm e}^{{\rm i}\theta}\rangle,
\end{equation}
where $|0\rangle$ and $|1\rangle$ represent the Fock states that
contain $0$ and $1$ photon, respectively. $|\alpha\rangle$
represents a coherent state. $\theta= \chi t$ represents a phase
shift with the interaction time $t$. With this cross-Kerr
nonlinearity, the HBSA protocol for the 16 polarization-spatial
hyperentangled Bell states can be implemented with two steps,
including the spatial-mode Bell-state analysis and the polarization
Bell-state analysis, shown in Figs.~\ref{HBSA1} and
 ~\ref{HBSA2}, respectively.

The setup of the spatial-mode Bell-state analysis is shown in
Fig.~\ref{HBSA1}, which is constructed with the spatial-mode
parity-check QNDs. After the two photons $A$ and $B$ pass through the
quantum circuit shown in Fig.~\ref{HBSA1}a in sequence, the state
of the quantum system composed of the two-photon system
$AB$ in the spatial-mode DOF and the coherent state $|\alpha\rangle$   evolves to
\begin{equation}  
\begin{split}
|\phi^\pm\rangle^{AB}_{S}|\alpha\rangle&\;\rightarrow\;
\frac{1}{\sqrt{2}}\left(|a_1b_1\rangle|\alpha
{\rm e}^{{\rm i}\theta}\rangle\pm|a_2b_2\rangle|\alpha
{\rm e}^{-{\rm i}\theta}\rangle\right),\\
|\psi^\pm\rangle^{AB}_{S}|\alpha\rangle&\;\rightarrow\;
\frac{1}{\sqrt{2}}\left(|a_1b_2\rangle|\alpha
\rangle\pm|a_2b_1\rangle|\alpha\rangle\right).
\end{split}
\end{equation}
Then the coherent beam is detected by an X-quadrature measurement, and the
states $|\alpha {\rm e}^{{\rm i}\theta}\rangle$ and $|\alpha {\rm e}^{-{\rm i}\theta}\rangle$
cannot be distinguished. Hence the Bell states $|\phi^\pm\rangle^{AB}_{S}$
can be distinguished from the Bell states $|\psi^\pm\rangle^{AB}_{S}$ with the
homodyne-heterodyne measurements. If the coherent state has a phase
shift $\theta$ ($-\theta$), the spatial-mode state of the two-photon
system $AB$ is one of the states $|\phi^\pm\rangle^{AB}_{S}$.
If the coherent state has no phase
shift, the spatial-mode state of the two-photon
system $AB$ is one of the states $|\psi^\pm\rangle^{AB}_{S}$.

Subsequently, the two photons $A$ and $B$ are put into the BS shown in
Fig.~\ref{HBSA1}b in sequence, and the spatial-mode state of the
two-photon system $AB$ is transformed into
\begin{equation}  
\begin{split}
|\phi^+\rangle^{AB}_S \;\;&\rightarrow\;\;|\phi^+\rangle^{AB}_S, \;\;\;\;
|\phi^-\rangle^{AB}_S \;\;\rightarrow\;\;|\psi^+\rangle^{AB}_S, \\
|\psi^+\rangle^{AB}_S \;\;&\rightarrow\;\;|\phi^-\rangle^{AB}_S, \;\;\;\;
|\psi^-\rangle^{AB}_S \;\;\rightarrow\;\; |\psi^-\rangle^{AB}_S .
 \end{split}
\end{equation}
After the two photons $AB$ and the coherent beam pass through the
cross-Kerr medium in Fig.~\ref{HBSA1}b, the four spatial-mode Bell states can be
distinguished by the X-quadrature measurement on the coherent beam.
If the coherent state has a phase shift $\theta_1+\theta_3$, $\theta_2+\theta_4$,
$\theta_1+\theta_4$, or $\theta_2+\theta_3$, the spatial mode of the  two-photon
system $AB$ is $c_1d_1$, $c_2d_2$, $c_1d_2$, or $c_2d_1$, respectively.
The initial spatial-mode state $|\phi^+\rangle^{AB}_S$ (or $|\psi^+\rangle^{AB}_S$)
corresponds to the spatial mode $c_1d_1$ or $c_2d_2$, and
the initial spatial-mode state $|\phi^-\rangle^{AB}_S$ (or $ |\psi^-\rangle^{AB}_S$)
corresponds to the spatial mode $c_1d_2$ or $c_2d_1$.

The setup of the polarization Bell-state analysis is shown in
Fig.~\ref{HBSA2}, which is constructed with the polarization
parity-check QND. After the two photons $A$ and $B$ pass through the
polarization parity-check QND shown in Fig.~\ref{HBSA2} in sequence,
the state of the system composed of the two-photon system $AB$ in
the polarization DOF and the coherent state $|\alpha\rangle$ evolves
to
\begin{equation}  
\begin{split}
|\phi^\pm\rangle^{AB}_{P}|\alpha\rangle&\rightarrow \frac{1}{\sqrt{2}}\left(|HH\rangle|\alpha {\rm e}^{{\rm i}\theta}\rangle\pm|VV\rangle|\alpha {\rm e}^{-{\rm i}\theta}\rangle\right), \\
|\psi^\pm\rangle^{AB}_{P}|\alpha\rangle&\rightarrow
\frac{1}{\sqrt{2}}\left(|HV\rangle|\alpha
\rangle\pm|VH\rangle|\alpha\rangle\right).
\end{split}
\end{equation}
After the X-quadrature measurement  is performed on the coherent
beam, the Bell states $|\phi^\pm\rangle^{AB}_{P}$ can be
distinguished from the Bell states $|\psi^\pm\rangle^{AB}_{P}$. If
the coherent state has a phase shift $\theta$ ($-\theta$), the
polarization state of the two-photon system $AB$ is one of the
states $|\phi^\pm\rangle^{AB}_{P}$. If the coherent state has no
phase shift, the polarization state of the two-photon system $AB$ is
one of the states $|\psi^\pm\rangle^{AB}_{P}$.

Subsequently, the two photons $A$ and $B$ are put into the H$_P$ shown
in Fig.~\ref{HBSA2}, and the polarization state of the two-photon system
$AB$ is transformed into
\begin{equation}  
\begin{split}
|\phi^+\rangle^{AB}_P \;\;&\rightarrow\;\;|\phi^+\rangle^{AB}_P, \;\;\;\;
|\phi^-\rangle^{AB}_P \;\;\rightarrow\;\;|\psi^+\rangle^{AB}_P, \\
|\psi^+\rangle^{AB}_P \;\;&\rightarrow\;\;|\phi^-\rangle^{AB}_P, \;\;\;\;
|\psi^-\rangle^{AB}_P \;\;\rightarrow\;\; |\psi^-\rangle^{AB}_P .
 \end{split}
\end{equation}
Then the four polarization Bell states can be distinguished by the
result of four single-photon detectors. If the detectors $H_a,V_b$
(or $V_a,H_b$) click, the initial polarization state is
$|\phi^-\rangle^{AB}_P$ (or $|\psi^-\rangle^{AB}_P$). If the
detectors $H_a,H_b$ (or $V_a,V_b$) click, the initial polarization
state is $|\phi^+\rangle^{AB}_P$ (or $|\psi^+\rangle^{AB}_P$). In
this way, one can completely distinguish  the 16 hyperentangled Bell
states by using the spatial-mode parity-check QNDs, polarization
parity-check QND, and single-photon detectors.

\subsection{Teleportation with a hyperentangled channel}

The quantum teleportation protocol is used to transfer the unknown
information of a quantum state between between the two remote users
\cite{QT1}. With hyperentanglement, two-qubit unknown information
can be transferred by teleporting a photon \cite{HBSA}.

\begin{center}
\begin{figure}[!h]
\includegraphics*[width=6cm]{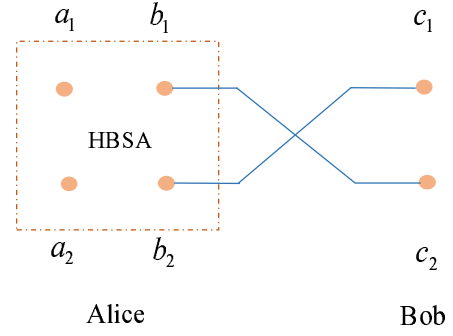}  
\caption{Schematic diagram of the quantum teleportation protocol
with hyperentanglement \cite{HBSA}.}\label{hypertele}
\end{figure}
\end{center}

The principle of quantum teleportation protocol with
hyperentanglement is shown in Fig.~\ref{hypertele}. The photon $A$ is
in the state
$|\varphi\rangle_A=(\alpha|H\rangle+\beta|V\rangle)_A\otimes(\gamma|a_1\rangle
+\delta|a_2\rangle)$, and the photon pair $BC$ is in a
hyperentangled Bell state
$|\phi^+\rangle_{BC}=\frac{1}{2}(|HH\rangle+|VV\rangle)_{BC}\otimes(|b_1c_1\rangle
+|b_2c_2\rangle)$, where the photons $B$ and $C$ are obtained by the
two remote users Alice and Bob, respectively. Alice can transfer the
two-qubit information of photon $A$ to Bob by performing HBSA on the
two photons $A$ and $B$.

The state of the three-photon system $ABC$ can be rewritten as
\begin{equation}  
\begin{split}
\!\!\!\!\!\!\!\!|\varphi&\rangle_A\otimes|\phi^+\rangle_{BC}\\
=&\frac{1}{4}\bigg\{\Big[|\phi^+\rangle_P^{AB}(\alpha|H\rangle+\beta|V\rangle)_c
+|\phi^-\rangle_P^{AB}(\alpha|H\rangle-\beta|V\rangle)_c\\
&+\!|\psi^+\rangle_P^{AB}(\alpha|V\rangle+\beta|H\rangle)_c
+|\psi^-\rangle_P^{AB}(\alpha|V\rangle-\beta|H\rangle)_c\Big]\\
&\otimes\!\Big[|\phi^+\rangle_S^{AB}(\gamma|c_1\rangle+\delta|c_2\rangle)
+|\phi^-\rangle_S^{AB}(\gamma|c_1\rangle-\delta|c_2\rangle)\\
&+\!|\psi^+\rangle_S^{AB}(\gamma|c_2\rangle+\delta|c_1\rangle)
+|\psi^-\rangle_S^{AB}(\gamma|c_2\rangle-\delta|c_1\rangle)\Big]\bigg\}.
 \end{split}
\end{equation}
After Alice performs HBSA on the two photons $A$ and $B$, photon $C$
will be projected to a single-photon quantum state in two DOFs. If
the outcome of HBSA for the photon pair $AB$ is
$|\phi^\pm\rangle_P|\phi^\pm\rangle_S$,
$|\psi^\pm\rangle_P|\phi^\pm\rangle_S$,
$|\phi^\pm\rangle_P|\psi^\pm\rangle_S$, or
$|\psi^\pm\rangle_P|\psi^\pm\rangle_S$, the state of photon $C$ is
projected to
$(\alpha|H\rangle\pm\beta|V\rangle)(\gamma|c_1\rangle\pm\delta|c_2\rangle)$,
$(\alpha|V\rangle\pm\beta|H\rangle)(\gamma|c_1\rangle\pm\delta|c_2\rangle)$,
$(\alpha|H\rangle\pm\beta|V\rangle)(\gamma|c_2\rangle\pm\delta|c_1\rangle)$,
or
$(\alpha|V\rangle\pm\beta|H\rangle)(\gamma|c_2\rangle\pm\delta|c_1\rangle)$,
respectively. If the polarization (spatial-mode) state of photon
pair $AB$ is $|\phi^-\rangle_P$ ($|\phi^-\rangle_S$), Bob should
perform a polarization (spatial-mode) phase-flip operation
$\sigma^P_z$ ($\sigma^S_z$) on photon $C$ after Alice publishes  the
result of HBSA. If the polarization (spatial-mode) state of photon
pair $AB$ is $|\psi^+\rangle_P$ ($|\psi^+\rangle_S$), Bob should
perform a polarization (spatial-mode) bit-flip operation
$\sigma^P_x$ ($\sigma^S_x$) on photon $C$. If the polarization
(spatial-mode) state of photon pair $AB$ is $|\psi^-\rangle_P$
($|\psi^-\rangle_S$), Bob should perform a unitary operation $-{\rm
i}\sigma^P_y$ ($-{\rm i}\sigma^S_y$) on photon $C$. Then, Bob can
obtain the unknown single-photon state
$(\alpha|H\rangle+\beta|V\rangle)(\gamma|c_1\rangle+\delta|c_2\rangle)$.
Here, $\sigma^P_z=|H\rangle\langle H|-|V\rangle\langle V|$,
$\sigma^P_x=|H\rangle\langle V|+|V\rangle\langle H|$, $-{\rm
i}\sigma^P_y=|H\rangle\langle V|-|V\rangle\langle H|$,
$\sigma^S_z=|c_1\rangle\langle c_1|-|c_2\rangle\langle c_2|$,
$\sigma^S_x=|c_1\rangle\langle c_2|+|c_2\rangle\langle c_1|$, and
$-{\rm i}\sigma^S_y=|c_1\rangle\langle c_2|-|c_2\rangle\langle
c_1|$.

\subsection{Hyperentanglement swapping}

\begin{center}
\begin{figure}[!h]
\includegraphics*[width=7cm]{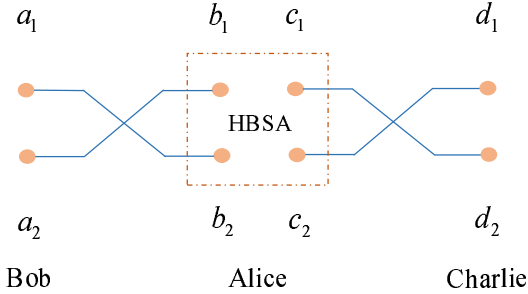}   
\caption{Schematic diagram of the hyperentanglement swapping
protocol \cite{HBSA,HBSA1}.}\label{hyperswap}
\end{figure}
\end{center}

Entanglement swapping is used to obtain the entanglement between two
particles that have no interaction initially, and it has been widely
applied in quantum repeaters and quantum communication protocols.
The principle of hyperentanglement swapping is shown in
Fig.~\ref{hyperswap}. The photon pairs $AB$ and $CD$ are initially
in the hyperentangled Bell states $|\phi^+\rangle^{AB}_{PS}$ and
$|\phi^+\rangle^{CD}_{PS}$, respectively. Here,
\begin{equation}                           \label{eq.13}   
\begin{split}
|\phi^+\rangle^{AB}_{PS} \; &= \; \frac{1}{2}(|HH\rangle\!+\!|VV\rangle)^{AB}_P\!\otimes\!(|a_1b_1\rangle\!+\!|a_2b_2\rangle)^{AB}_S, \\
|\phi^+\rangle^{CD}_{PS} \; &= \;
\frac{1}{2}(|HH\rangle\!+\!|VV\rangle)^{CD}_P\!\otimes\!(|c_1d_1\rangle\!+\!|c_2d_2\rangle)^{CD}_S.
 \end{split}
\end{equation}
The photons $B$ and $C$   belong to Alice. The photons $A$ and
$D$  belong  to Bob and Charlie, respectively. That is, Alice shares
a hyperentangled photon pair with Bob, and she also shares a
hyperentangled photon pair with Charlie. The task of
hyperentanglement swapping is to obtain the hyperentangled Bell
state
$|\phi^+\rangle^{AD}_{PS}=\frac{1}{2}(|HH\rangle+|VV\rangle)^{AD}_P\otimes(|a_1d_1\rangle
+|a_2d_2\rangle)^{AD}_S$, which can be implemented by performing
HBSA on photon pair $BC$.

The state of the four-photon system $ABCD$
can be rewritten as
\begin{equation}                           \label{eq.14}    
\begin{split}
|\phi^+\rangle^{AB}_{PS}\otimes&|\phi^+\rangle^{CD}_{PS}\\
=&\frac{1}{4}\Big[\big(|\phi^+\rangle^{AD}_P|\phi^+\rangle^{BC}_P
+|\phi^-\rangle^{AD}_P|\phi^-\rangle^{BC}_P\\
&+|\psi^+\rangle^{AD}_P|\psi^+\rangle^{BC}_P + |\psi^-\rangle^{AD}_P|\psi^-\rangle^{BC}_P\big)\\
&\otimes\big(|\phi^+\rangle^{AD}_S|\phi^+\rangle^{BC}_S
+|\phi^-\rangle^{AD}_S|\phi^-\rangle^{BC}_S\\
&+|\psi^+\rangle^{AD}_S|\psi^+\rangle^{BC}_S
+|\psi^-\rangle^{AD}_S|\psi^-\rangle^{BC}_S\big)\Big].
 \end{split}
\end{equation}
After Alice performs HBSA on the photon pair $BC$, the correlation
between the two photons $AD$ can be created. If the outcome of HBSA
for the photon pair $BC$ is $|\phi^\pm\rangle_P|\phi^\pm\rangle_S$,
$|\psi^\pm\rangle_P|\phi^\pm\rangle_S$,
$|\phi^\pm\rangle_P|\psi^\pm\rangle_S$, or
$|\psi^\pm\rangle_P|\psi^\pm\rangle_S$, the state of the photon pair
$AD$ is projected to $|\phi^\pm\rangle_P|\phi^\pm\rangle_S$,
$|\psi^\pm\rangle_P|\phi^\pm\rangle_S$,
$|\phi^\pm\rangle_P|\psi^\pm\rangle_S$, or
$|\psi^\pm\rangle_P|\psi^\pm\rangle_S$, respectively. If the
polarization (spatial-mode) state of the photon pair $BC$ is
$|\phi^-\rangle_P$ ($|\phi^-\rangle_S$), Bob should perform a
unitary operation $\sigma^P_z$ ($\sigma^S_z$) on photon $A$ after
Alice publishes the result of HBSA. If the polarization
(spatial-mode) state of the photon pair $BC$ is $|\psi^+\rangle_P$
($|\psi^+\rangle_S$), Bob should perform a unitary operation
$\sigma^P_x$ ($\sigma^S_x$) on photon $A$. If the polarization
(spatial-mode) state of the photon pair $BC$ is $|\psi^-\rangle_P$
($|\psi^-\rangle_S$), Bob should perform a unitary operation
$-{\rm i}\sigma^P_y$ ($-{\rm i}\sigma^S_y$) on photon $A$. Now, Bob and Charlie
can share a hyperentangled photon pair $AD$ in the state
$|\phi^+\rangle^{AD}_{PS}$.

\section{Hyperentanglement concentration}

\subsection{Development of entanglement concentration}

In the practical quantum communication with entanglement,  the
entangled photon systems are produced locally,  which leads to their
decoherence when the photons are transmitted over a quantum channel
with environment noise or stored in practical quantum devices.
Quantum repeater is a necessary technique for long-distance quantum
communication and it is used to overcome the influence from this
decoherence \cite{repeater}. In fact, the optimal way to overcome
the influence on photon systems from channel noise in quantum
communication is the self-error-rejecting qubit transmission
\cite{LIXHAPL} with linear optics as it is an active way to decrease
the influence from channel noise and it is very  efficient and
simple to be implemented in experiment with current feasible
techniques. However, this scheme \cite{LIXHAPL} can only depress
most of the influence from the channel noise in the process of
photon distribution, as the same as the other active methods for
overcoming the influence from noise \cite{DFSa1,DFSa2,DFSa3}. It
does not work in depressing the influence of noise from both a
long-distance channel and the storage process for quantum states.
Moreover, quantum repeaters for long-distance quantum communication
require the entangled photons with higher fidelity (usually $\sim$
99\%) beyond  that from faithful qubit transmission schemes (about
90\% $\sim$ 96\% for a polarization quantum state of photons over an
optical-fiber channel with several kilometers). That is,
entanglement concentration and entanglement purification are not
only useful but also absolutely necessary in long-distance quantum
communication.

Entanglement concentration is used to distill some nonlocal
entangled systems in a maximally entangled state from a set of
nonlocal entangled systems in a partially entangled pure state
\cite{ECP1}. Before 2013, entanglement concentration is focused on
the nonlocal quantum  states in one DOF, such as the polarization
states of photons, the two-level quantum states of atom systems, or
the spins of electron systems.  The first entanglement concentration
protocol (ECP) was proposed by Bennett et al. \cite{ECP1} in 1996,
which is based on the Schmidt projection \cite{ECP1}. Also, it is
just a mathematic method for entanglement concentration.  In 2001,
two ECPs were proposed \cite{ECP5,ECP4} with PBSs for two ideal
entangled photon sources. In 2008, Sheng et al. \cite{ECP6} proposed
a repeatable ECP  to concentrate both bipartite and multipartite
quantum systems, and it has an advantage of far higher efficiency
and yield than those in  Bennett's ECP \cite{ECP1} and the PBS-based
ECPs \cite{ECP5,ECP4}, by iteration of the concentration process two
or three times. In fact, depending on whether the parameters of the
nonlocal less-entangled states are unknown
\cite{ECP1,ECP5,ECP4,ECP6,ECPeletronWangC} or known
\cite{ECP2,ECPShi,ECP7}, the existing ECPs can be classed into two
groups. When the parameters are known, one nonlocal photon system is
enough for concentrating the nonlocal entanglement efficiently
\cite{ECP2,ECPShi,ECP7} with far higher yield than those with
unknown parameters  \cite{ECP1,ECP5,ECP4,ECP6}. In 1999, Bose et al.
\cite{ECP2} designed the first ECP for nonlocal entangled photon
pairs in the less-entangled pure state with known parameters,
resorting to the entanglement swapping of a nonlocal entangled
photon pair and a local entangled photon pair. In 2000, Shi et al.
\cite{ECPShi} proposed another ECP based on entanglement swapping
and a collective unitary operation on two qubits. In 2012, Sheng et
al.  \cite{ECP7} presented two ECPs for photon systems in the
less-entangled states with known parameters, according to which an
ancillary single photon state can be prepared to assist the
concentration. In 2012, Deng  \cite{ECP8} presented the optimal
nonlocal multipartite ECP based on projection measurement. Also,
some schemes for concentrating W states have been proposed
\cite{ecpadd1,ecpadd2,ecpadd3}.  Moreover, two groups
\cite{ECP51,ECP41} demonstrated in experiment the entanglement
concentration of two-photon systems with linear optical elements. A
review about entanglement concentration of photon systems in one DOF
with cross-Kerr nonlinearity was presented in Ref.~\cite{ECPreview}.

The investigation on hyperentanglement concentration began from
2013. In this year, Ren et al. \cite{HECP} proposed the
parameter-splitting method to extract the maximally entangled
photons in both the polarization and spatial-mode DOFs when the
coefficients of the initial partially hyperentangled states are
known. This fascinating (novel) method is very efficient and simple
in terms of concentrating partially entangled states as it can be
achieved with the maximal success probability by performing the
protocol only once, resorting to linear-optical elements only, not
nonlinearity, no matter what the form of the known nonlocal
entangled state is, what the number of the DOFs is, and what the
number of particles in the quantum system is. They \cite{HECP} also
gave the first hyperentanglement concentration protocol (hyper-ECP)
for the unknown polarization-spatial less-hyperentangled states with
linear-optical elements only and another hyper-ECP \cite{HEPPECP}
for nonlocal polarization-spatial less-hyperentangled states with
unknown parameters assisted by diamond nitrogen vacancy (NV) centers
inside photonic crystal cavities. Subsequently, Ren and Long
\cite{HECP1} proposed a general hyper-ECP for photon systems
assisted by quantum-dot spins inside optical microcavities and
another high-efficiency hyper-ECP \cite{HECP3} with the
quantum-state-joining method. In the same time, Li and Ghose
\cite{HECP2} presented a hyper-ECP resorting to linear optics. In
2015, they also brought forward an efficient hyper-ECP for the
multipartite hyperentangled state via the cross-Kerr nonlinearity
\cite{HECPLixhOE} and another hyper-ECP for time-bin and
polarization hyperentangled photons \cite{HECP4}. In 2016, Cao et
al. \cite{HECPadd2} presented a hyper-ECP for entangled photons by
using photonic module system.

In this section, we overview the hyper-ECPs for high-capacity
long-distance quantum communication \cite{HECP}, resorting to the
parameter-splitting method  \cite{HECP} and the Schmidt projection
method \cite{ECP1}, respectively. With the parameter-splitting
method  \cite{HECP}, the hyper-ECP can be implemented with the
maximal success probability \cite{HECP}, resorting to linear optical
elements only. With the Schmidt projection method, the success
probability of the hyper-ECP is relatively low with linear optical
elements \cite{HECP,HECP2,HECP4}, and it can be improved by
iterative application of the hyper-ECP process with nonlinear
optical elements \cite{HECP1,HEPPECP}.

\subsection{Hyper-ECP with parameter-splitting method}
\label{sec3.1}

The parameter-splitting method is introduced to concentrate nonlocal
partially entangled states with their parameters accurately
known to the remote users \cite{HECP}. With this method, only one remote user
has to perform local operations with linear optical elements, and the
success probability of the ECP can achieve the maximal value. The
ECP for polarization (spatial-mode) DOF of photon system is introduced
in detail in Ref.~\cite{HECP}. Here, we introduce the hyper-ECP for polarization-spatial
hyperentangled Bell state by using the parameter-splitting method \cite{HECP}. That is,
Alice and Bob obtain a subset of nonlocal two-photon systems in a maximally
hyperentangled Bell state by splitting the parameters of the
partially hyperentangled Bell states with linear optical elements only.

The partially hyperentangled Bell state is described as
\begin{eqnarray}                            
|\phi_0\rangle_{AB}=(\alpha|HH\rangle+\beta|VV\rangle)_{AB}
\otimes(\gamma|a_1b_1\rangle+\delta|a_2b_2\rangle),\nonumber\\
\end{eqnarray}
where the subscripts $A$ and $B$ represent two photons obtained by
the two remote users, Alice and Bob, respectively. $\alpha$,
$\beta$, $\gamma$, and $\delta$ are four real parameters that are
known to the two remote users, and they satisfy the relation
$|\alpha|^2+|\beta|^2=|\gamma|^2+|\delta|^2=1$.

\begin{center}                      
\begin{figure}[!h]
\includegraphics*[width=7.5cm]{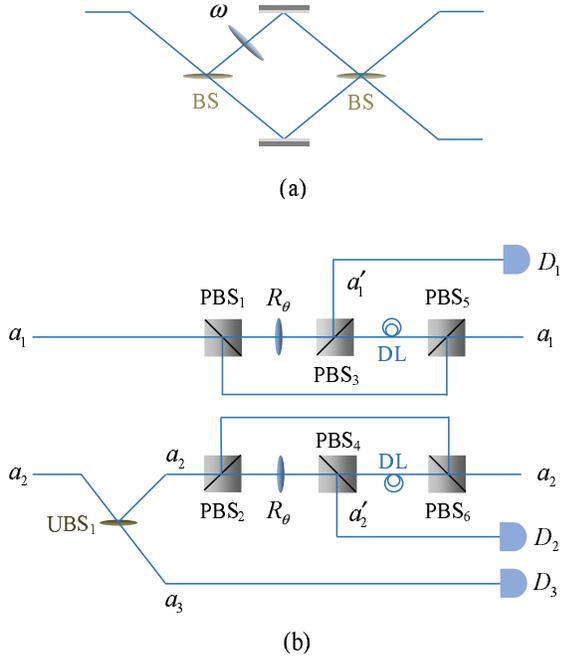}
\caption{(a) Schematic diagram of an unbalanced BS (UBS)
\cite{HECP}. $\omega$ represents a wave plate which can cause a
phase shift between the two spatial modes. (b) Schematic diagram of
the polarization-spatial hyper-ECP with the parameter-splitting
method  \cite{HECP}. UBS represents an unbalanced beam splitter with
the reflection coefficient $R=\gamma/\delta$. $R_\theta$ represents
a wave plate which is used to rotate the horizontal polarization
with an angle $\theta=arccos(\beta/\alpha)$. DL denotes a time-delay
device which is used to make the two wavepackets of the two spatial
modes arrive at PBS$_5$ (or PBS$_6$) in the same time. $D_i$
($i=1,2,3$) represents a single-photon detector.}\label{figure3.1}
\end{figure}
\end{center}

The setup of the hyper-ECP \cite{HECP} for the partially hyperentangled Bell
state $|\phi_0\rangle_{AB}$ is shown in Fig.~\ref{figure3.1}b.
It is implemented by performing some local unitary operations on
both the spatial-mode and polarization DOFs of photon $A$. No
operation is performed on photon $B$. To describe
the principle of the hyper-ECP explicitly and simply, the four parameters are
chosen as $|\alpha| > |\beta|$ and $|\gamma| < |\delta|$. In other cases, the hyper-ECP
can be implemented as the same as this one with or
without a little modification.

First, Alice splits the parameter of the spatial-mode state by
performing a unitary operation on spatial mode $a_2$, resorting to
an unbalanced beam splitter (i.e., UBS) with reflection coefficient
$R=\gamma/\delta$ (shown in Fig.~\ref{figure3.1}a). The state of the
photon pair $AB$ is changed from $|\phi_0\rangle_{AB}$ to
$|\phi_1\rangle_{AB}$. Here
\begin{eqnarray}                             
|\phi_1\rangle_{AB}\!&=&\!(\alpha|HH\rangle+\beta|VV\rangle)_{AB}
\otimes\Big[\gamma(|a_1b_1\rangle
+|a_2b_2\rangle)\nonumber\\
\!&&+\sqrt{|\delta|^2-|\gamma|^2}|a_3b_2\rangle\Big].
\end{eqnarray}
If photon $A$ is not detected in the spatial mode $a_3$, the
spatial-mode state of the photon pair $AB$ is transformed into a
maximally entangled Bell state.

Subsequently, Alice splits the parameter of the polarization state by performing
the same polarization unitary operations on the spatial modes $a_1$ and
$a_2$ as shown in Fig.~\ref{figure3.1}b. After two spatial modes
$a_1$ and $a_2$ pass through PBSs (i.e., PBS$_1$ and PBS$_2$) and
$R_\theta$, the state of the photon pair $AB$ is transformed from
$|\phi_1\rangle_{AB}$ to $|\phi_2\rangle_{AB}$. Here
\begin{eqnarray}                            
|\phi_2\rangle_{AB}\!\!&=&\!\!\left[\beta(|HH\rangle+|VV\rangle)
+\sqrt{|\alpha|^2-|\beta|^2}|V'H\rangle\right]_{AB}
\nonumber\\
&&\otimes\gamma(|a_1b_1\rangle+|a_2b_2\rangle)+(\alpha|HH\rangle+\beta|VV\rangle)_{AB}
\nonumber\\
&&\otimes\sqrt{|\delta|^2-|\gamma|^2}\,|a_3b_2\rangle,
\end{eqnarray}
where $\vert V'\rangle$ represents the vertical polarization of photon
$A$ after an operation $R_\theta$. The wave plate $R_\theta$ is used
to perform a rotate operation $\vert H\rangle\rightarrow {\rm cos}\theta
\vert H\rangle +  {\rm sin}\theta \vert V\rangle$ on the horizontal
polarization $\vert H\rangle$.

Finally, Alice lets two spatial modes $a_1$ and $a_2$ pass through PBS$_3$, PBS$_4$,
DL, PBS$_5$ and PBS$_6$, and the state of the photon pair $AB$ is transformed
from $|\phi_2\rangle_{AB}$ to $|\phi_3\rangle_{AB}$. Here
\begin{eqnarray}                            
|\phi_3\rangle_{AB}\!&=&\!\beta\gamma(|HH\rangle+|VV\rangle)_{AB}(|a_1b_1\rangle
+|a_2b_2\rangle)
\nonumber\\
\!&&+\gamma\sqrt{|\alpha|^2-|\beta|^2}|VH\rangle_{AB}(|a'_1b_1\rangle
+|a'_2b_2\rangle)
\nonumber\\
\!&&+\sqrt{|\delta|^2-|\gamma|^2}(\alpha|HH\rangle
+\beta|VV\rangle)_{AB}|a_3b_2\rangle.\nonumber\\
\label{HBSoutcome}
\end{eqnarray}
If photon $A$ is not detected in one of the spatial modes $a'_1$ and $a'_2$, the
polarization state of the photon pair $AB$ is transformed into a
maximally entangled Bell state. That is, the maximally
hyperentangled Bell state $|\phi\rangle_{AB}$ is obtained. Here
\begin{eqnarray}                           
|\phi\rangle_{AB} = \frac{1}{2}(|HH\rangle+|VV\rangle)_{AB}
(|a_1b_1\rangle+|a_2b_2\rangle).\;\;\;\;
\end{eqnarray}
If photon $A$ is detected in one of the spatial modes $a'_1$, $a'_2$, and $a_3$,
the polarization DOF or the spatial-mode DOF of the photon pair $AB$
will project to a product state, which means the hyper-ECP fails.
According to the detection in the spatial modes of photon $A$, Alice
can read out whether the hyper-ECP succeeds or not in theory. As the
efficiency of a single-photon detector is lower than 100 \%, the
mistaken of a successful event caused by the detection inefficiency can
be eliminated by postselection.

The success probability of this hyper-ECP is $P=4|\beta\gamma|^2$,
which achieves the maximal success probability for obtaining a
maximally hyperentangled Bell state from a partially hyperentangled
Bell state. Moreover, this parameter-splitting method is suitable
for all the entanglement concentration of photon systems in nonlocal
partially entangled pure states with known parameters, including
those based on one DOF and those based on multiple DOFs.


\subsection{Hyper-ECP with Schmidt projection method} \label{sec3.2}

Here, we mainly introduce two hyper-ECPs for polarization-spatial
hyperentangled Bell states with unknown parameters \cite{HECP}. The
first one is implemented with linear optical elements \cite{HECP},
which is much easier to achieve in experiment. The second one is
implemented with nonlinear optical elements \cite{HEPPECP}, which
can improve the success probability by iterative application of the
hyper-ECP.

\subsubsection{Hyper-ECP with linear optical elements}
\label{sec3.2.1}

In the Schmidt projection method, two identical photon pairs $AB$
and $CD$ are required, which are in the partially hyperentangled
Bell states $|\phi_0\rangle_{AB}$ and $|\phi_0\rangle_{CD}$,
respectively. Here
\begin{eqnarray}                           
|\phi_0\rangle_{AB}\!&=&\!(\alpha|HH\rangle+\beta|VV\rangle)_{AB}
\otimes(\gamma|a_1b_1\rangle+\delta|a_2b_2\rangle),\nonumber\\
|\phi_0\rangle_{CD}\!&=&\!(\alpha|HH\rangle+\beta|VV\rangle)_{CD}
\otimes(\gamma|c_1d_1\rangle+\delta|c_2d_2\rangle).\nonumber\\
\end{eqnarray}
Here the subscripts $AB$ and $CD$ represent two photon pairs shared
by the two remote users. Alice has the two photons $A$ and $C$, and
Bob has the two photons $B$ and $D$. $\alpha$, $\beta$, $\gamma$,
and $\delta$ are four unknown real parameters, and they satisfy the
relation $|\alpha|^2+|\beta|^2=|\gamma|^2+|\delta|^2=1$.

\begin{center}                         
\begin{figure}[!h]
\includegraphics*[width=6.4cm]{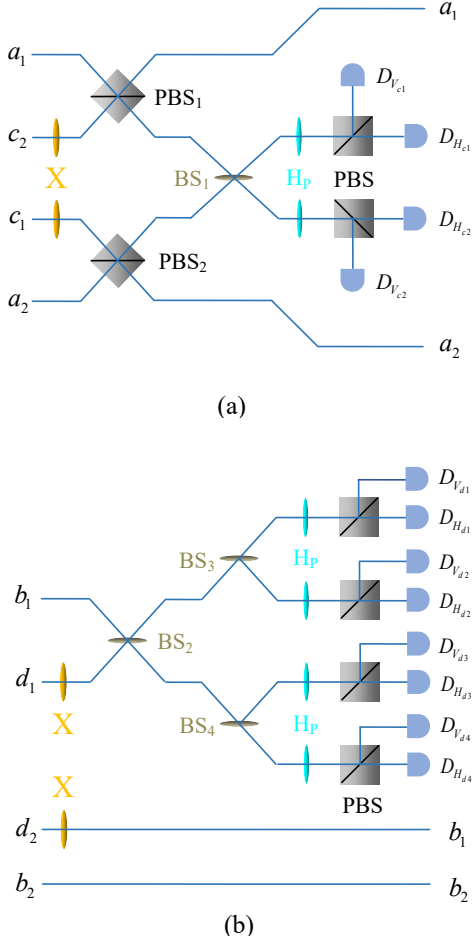}
\caption{Schematic diagram of the polarization-spatial hyper-ECP for
partially hyperentangled Bell states with unknown parameters
\cite{HECP}, resorting to the Schmidt projection method. (a)
Operations performed by Alice. (b) Operations performed by Bob. X
represents a half-wave plate which is used to perform a polarization
bit-flip operation $\sigma^P_x=|H\rangle\langle V| +
|V\rangle\langle H|$.}\label{figure3.2}
\end{figure}
\end{center}

The setup of the hyper-ECP with the Schmidt projection method for unknown
partially hyperentangled Bell states \cite{HECP}  is
shown in Fig.~\ref{figure3.2}. The initial state of four-photon
system $ABCD$ is described as:
$|\Phi_0\rangle=|\phi_0\rangle_{AB}\otimes|\phi_0\rangle_{CD}$.
After bit-flip operations ($\sigma^P_x$) are performed on the polarization DOF of
photons $C$ and $D$, the state of four-photon system $ABCD$ becomes
\begin{eqnarray}                          
|\Phi_1\rangle\!\!&=&\!\! (\alpha^2|HHVV\rangle +
\alpha\beta|VVVV\rangle
 + \alpha\beta|HHHH\rangle \nonumber\\
\!\!\!\!&& + \beta^2|VVHH\rangle)_{ABCD}
\otimes(\gamma^2|a_1b_1c_1d_1\rangle\nonumber\\
\!\!\!\!&& +\gamma\delta|a_2b_2c_1d_1\rangle+
\gamma\delta|a_1b_1c_2d_2\rangle+\delta^2|a_2b_2c_2d_2\rangle). \nonumber\\
\label{eq14}
\end{eqnarray}

Subsequently, the wavepackets from the spatial modes $a_1$ and
$c_2$ are put into PBS$_1$, and the wavepackets from the spatial
modes $a_2$ and $c_1$ are put into PBS$_2$. The wavepackets from
the spatial modes $b_1$ and $d_1$  are put into BS$_2$. Here PBSs in
Fig.~\ref{figure3.2}a are used to perform a polarization
parity-check measurement on the two photons $A$ and $C$, and BS in
Fig.~\ref{figure3.2}b is used to perform a spatial-mode
parity-check measurement on the two photons $B$ and $D$ with the
Hong-Ou-Mandel (HOM) effect \cite{HOM}. If the photon pair $AC$ is
in an even-parity polarization mode (i.e. $\vert HH\rangle_{AC}$ and
$\vert VV\rangle_{AC}$), only one photon will be detected by Alice
in principle, as shown in Fig.~\ref{figure3.2}a. If the photon
pair $AC$ is in an odd-parity polarization mode (i.e. $\vert
HV\rangle_{AC}$ and $\vert VH\rangle_{AC}$), both of the two photons
$A$ and $C$ will be detected or undetected in principle. If the
photon pair $BD$ is in an odd-parity spatial mode (i.e. $\vert b_1
d_2\rangle$ and $\vert b_2d_1\rangle$), only one photon will be
detected by Bob in principle as shown in Fig.~\ref{figure3.2}b. If
the photon pair $BD$ is in an even-parity spatial mode (i.e. $\vert
b_1d_1\rangle$ and $\vert b_2d_2\rangle$), both of the two photons
$B$ and $D$ will be detected or undetected in principle.

With the polarization parity-check measurement and spatial-mode
parity-check measurement, Alice and Bob can divide the polarization
states and the spatial-mode states of the four-photon systems into
two groups, respectively. They pick up the even-parity polarization
states of the photon pair $AC$ and the odd-parity spatial-mode states of the
photon pair $BD$, which leads to the fact that both Alice and Bob have
only one detector clicked. In this time, the state of the four-photon
system $ABCD$ is projected into the state $|\Phi_2\rangle$. Here
\begin{eqnarray}                           
|\Phi_2\rangle\!&=&\!\frac{1}{2}(|VVVV\rangle+|HHHH\rangle)_{ABCD}\nonumber\\
&&\otimes (|a_2b_2c_1d_1\rangle+|a_1b_1c_2d_2\rangle). \label{eq15}
\end{eqnarray}
If the outcome of the detectors is in another condition, this
hyper-ECP fails.

Finally, the Hadamard operations are performed on the
spatial-mode and polarization DOFs of the photons $C$ and $D$, respectively,
and the state of the four-photon system $ABCD$ is transformed from $|\Phi_2\rangle$
to $|\Phi_3\rangle$. Here
\begin{eqnarray}                           
|\Phi_3\rangle\!&=&\!\frac{1}{8}\big[(|VV\rangle+|HH\rangle)_{AB}(|VV\rangle+|HH\rangle)_{CD}\nonumber\\
&&+(|HH\rangle-|VV\rangle)_{AB}(|HV\rangle+|VH\rangle)_{CD}\big]\nonumber\\
&&\otimes\big[(|a_2b_2\rangle+|a_1b_1\rangle)(|c_1d_1\rangle+|c_2d_2\rangle)
\nonumber\\
&&-(|a_1b_1\rangle-|a_2b_2\rangle)(|c_1d_2\rangle+|c_2d_1\rangle)\big].
\end{eqnarray}
If the outcome of the two clicked detectors is in an even-parity
polarization mode and an even-parity spatial mode, the state of the
photon pair $AB$ is projected to the maximally hyperentangled Bell
state $|\phi\rangle_{AB}=\frac{1}{2}(|HH\rangle+|VV\rangle)_{AB}
(|a_1b_1\rangle+|a_2b_2\rangle)$. If the outcome of the two clicked
detectors is in an odd-parity polarization (spatial) mode, a phase-flip
operation $\sigma^P_z$
($\sigma^S_z$) on
photon $B$ is required to obtain the state $|\phi\rangle_{AB}$.

In principle, if Alice and Bob both have only one detector clicked,
the maximally hyperentangled Bell states can be obtained with the
probability of $P=4|\alpha\beta\gamma\delta|^2$. Otherwise, the
hyper-ECP fails. In practical, whether this hyper-ECP succeeds or
not can also be read out by postselection, if the efficiencies of
the single-photon detectors are lower than 100 \%.
With this method, the maximally hyperentangled GHZ states can be
obtained from the partially hyperentangled GHZ states \cite{HECP2} in the same way.

\subsubsection{Hyper-ECP with nonlinear optical elements} \label{sec3.1.3}

Here, we introduce the hyper-ECP with nonlinear optical elements.
That is, the polarization parity-check measurement and the
spatial-mode parity-check measurement are replaced by the
polarization parity-check QND (P-QND) and
the spatial-mode parity-check QND (S-QND), respectively, where the P-QND and S-QND
are constructed with the cross-Kerr nonlinearity.

\begin{center}                         
\begin{figure}[!h]
\includegraphics*[width=7.2cm]{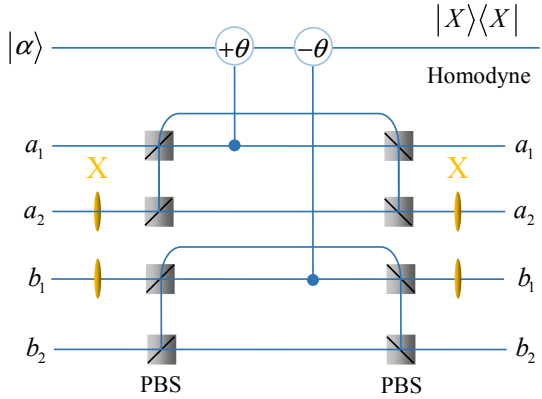}
\caption{Schematic diagram of the polarization parity-check quantum
nondemolition detector (P-QND) with the cross-Kerr nonlinearity
\cite{HECPLixhOE}.}\label{figure3.3}
\end{figure}
\end{center}

\emph{P-QND} --- The setup of the P-QND is shown in Fig.~\ref{figure3.3},
which is different from the one in Fig.~\ref{HBSA2}. In Fig.~\ref{HBSA2}, the
spatial-mode state is detected at the end of the spatial-mode Bell-state analysis.
Here, the P-QND is required to distinguish the even-parity polarization modes
from the odd-parity polarization modes without affecting the spatial-mode states.
That is,
\begin{eqnarray}                           
&&(\alpha|HH\rangle\pm\beta|VV\rangle)\otimes(\gamma|a_1b_1\rangle\pm\delta|a_2b_2\rangle)_{AB}|\alpha\rangle\nonumber\\
&& \rightarrow (\alpha|HH\rangle|\alpha {\rm e}^{{\rm i}\theta}\rangle\pm\beta|VV\rangle|\alpha {\rm e}^{-{\rm i}\theta}\rangle)_{AB}\nonumber\\
&& \;\;\;\otimes(\gamma|a_1b_1\rangle\pm\delta|a_2b_2\rangle), \nonumber\\
&&(\alpha|HH\rangle\pm\beta|VV\rangle)\otimes(\gamma|a_1b_2\rangle\pm\delta|a_2b_1\rangle)_{AB}|\alpha\rangle\nonumber\\
&& \rightarrow (\alpha|HH\rangle|\alpha {\rm e}^{{\rm i}\theta}\rangle\pm\beta|VV\rangle|\alpha {\rm e}^{-{\rm i}\theta}\rangle)_{AB}\nonumber\\
&& \;\;\;\otimes(\gamma|a_1b_2\rangle\pm\delta|a_2b_1\rangle), \nonumber\\
&&(\alpha|HV\rangle\pm\beta|VH\rangle)\otimes(\gamma|a_1b_1\rangle\pm\delta|a_2b_2\rangle)_{AB}|\alpha\rangle\nonumber\\
&& \rightarrow (\alpha|HV\rangle|\alpha\rangle\pm\beta|VH\rangle|\alpha\rangle)_{AB}\nonumber\\
&& \;\;\;\otimes(\gamma|a_1b_1\rangle\pm\delta|a_2b_2\rangle), \nonumber\\
&&(\alpha|HV\rangle\pm\beta|VH\rangle)\otimes(\gamma|a_1b_2\rangle\pm\delta|a_2b_1\rangle)_{AB}|\alpha\rangle\nonumber\\
&& \rightarrow (\alpha|HV\rangle|\alpha\rangle\pm\beta|VH\rangle|\alpha\rangle)_{AB}\nonumber\\
&& \;\;\;\otimes(\gamma|a_1b_2\rangle\pm\delta|a_2b_1\rangle).
 \;\;\;\;\label{eq6}
\end{eqnarray}
After the X-quadrature measurement is performed on the coherent beam, the
even-parity polarization Bell states can be distinguished from the
odd-parity polarization Bell states. If the coherent state has a
phase shift $\theta$ ($-\theta$), the polarization state of the
two-photon system $AB$ is in an even-parity mode. If the coherent
state has no phase shift,  the polarization state of the two-photon
system $AB$ is in the odd-parity mode.

\emph{S-QND} --- The setup of the S-QND in hyper-ECP is the same as the one in
Fig.~\ref{HBSA2}a. That is,
\begin{eqnarray}                           
&&(\alpha|HH\rangle\pm\beta|VV\rangle)
\otimes(\gamma|a_1b_1\rangle\pm\delta|a_2b_2\rangle)_{AB}|\alpha\rangle\nonumber\\
&& \rightarrow (\alpha|HH\rangle\rangle\pm\beta|VV\rangle)_{AB}\nonumber\\
&& \;\;\;\otimes(\gamma|a_1b_1\rangle|\alpha {\rm e}^{{\rm i}\theta}\rangle\pm\delta|a_2b_2\rangle|\alpha {\rm e}^{-{\rm i}\theta}\rangle), \nonumber\\
&&(\alpha|HH\rangle\pm\beta|VV\rangle)\otimes(\gamma|a_1b_2\rangle\pm\delta|a_2b_1\rangle)_{AB}|\alpha\rangle\nonumber\\
&& \rightarrow (\alpha|HH\rangle\pm\beta|VV\rangle)_{AB}\otimes(\gamma|a_1b_2\rangle|\alpha\rangle\pm\delta|a_2b_1\rangle|\alpha\rangle), \nonumber\\
&&(\alpha|HV\rangle\pm\beta|VH\rangle)\otimes(\gamma|a_1b_1\rangle\pm\delta|a_2b_2\rangle)_{AB}|\alpha\rangle\nonumber\\
&& \rightarrow (\alpha|HV\rangle\pm\beta|VH\rangle)_{AB}\nonumber\\
&& \;\;\;\otimes(\gamma|a_1b_1\rangle|\alpha {\rm e}^{{\rm i}\theta}\rangle\pm\delta|a_2b_2\rangle|\alpha {\rm e}^{-{\rm i}\theta}\rangle), \nonumber\\
&&(\alpha|HV\rangle\pm\beta|VH\rangle)\otimes(\gamma|a_1b_2\rangle\pm\delta|a_2b_1\rangle)_{AB}|\alpha\rangle\nonumber\\
&& \rightarrow
(\alpha|HV\rangle\pm\beta|VH\rangle)_{AB}\otimes(\gamma|a_1b_2\rangle|\alpha\rangle\pm\delta|a_2b_1\rangle|\alpha\rangle).\nonumber\\
\label{eq6}
\end{eqnarray}
After the X-quadrature measurement is performed on the coherent beam, the
even-parity spatial-mode Bell states can be distinguished from the
odd-parity spatial-mode Bell states. If the coherent state has a
phase shift $\theta$ ($-\theta$), the spatial-mode state of
two-photon system $AB$ is in an even-parity mode. If the coherent
state has no phase shift, the spatial-mode state of two-photon
system $AB$ is in an odd-parity mode.

\begin{figure*}
\includegraphics*[width=12cm]{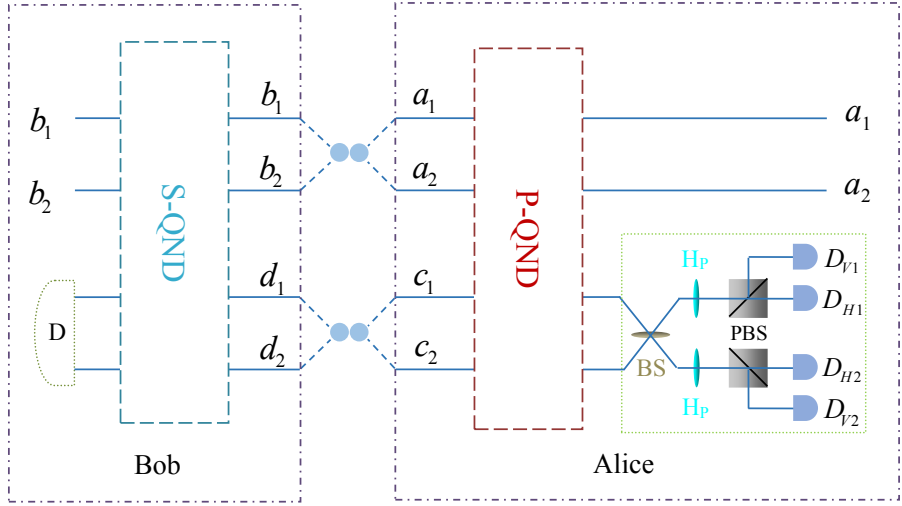}
\caption{Schematic diagram of the polarization-spatial hyper-ECP for
unknown partially hyperentangled Bell states
\cite{HECP1,HECPLixhOE}, resorting to the Schmidt projection method.
D represents the same operations as the ones performed by Alice in
the green dotted box. }\label{figure3.4}
\end{figure*}

In long-distance quantum communication, the maximally
hyperentangled Bell state $|\phi\rangle_{AB}$ may decay to the
partially hyperentangled Bell state $|\phi_0\rangle_{AB}$ by
the channel noise. Here
\begin{eqnarray}                           
|\phi\rangle_{AB}   \!&=&\! \frac{1}{2}(|HH\rangle+|VV\rangle)_{AB}\otimes(|a_1b_1\rangle+|a_2b_2\rangle),\nonumber\\
|\phi_0\rangle_{AB} \!&=&\! (\alpha|HH\rangle+\beta|VV\rangle)_{AB}\otimes(\gamma|a_1b_1\rangle+\delta|a_2b_2\rangle).\nonumber\\
\end{eqnarray}
Here, the two photons $A$ and $B$ are obtained by Alice and Bob, respectively.
$\alpha$, $\beta$, $\gamma$, and $\delta$ are four unknown real parameters and they satisfy the relation $|\alpha|^2+|\beta|^2=|\gamma|^2+|\delta|^2=1$.
The quantum circuit of the hyper-ECP for the partially hyperentangled Bell state $|\phi_0\rangle_{AB}$ is shown in Fig.~\ref{figure3.4}.
Two identical two-photon systems $AB$ and $CD$ are required in this hyper-ECP.
Here, $|\phi_0\rangle_{CD}=(\alpha|HH\rangle+\beta|VV\rangle)_{CD}
\otimes(\gamma|c_1d_1\rangle+\delta|c_2d_2\rangle)$,
and the two photons $C$ and $D$ are obtained by Alice and Bob, respectively.

\emph{The first round of the hyper-ECP} --- The initial state of the four-photon system $ABCD$ is $|\Phi_0\rangle=|\phi_0\rangle_{AB}\otimes|\phi_0\rangle_{CD}$.
Alice and Bob can divide the states of the four-photon systems
into two groups in the polarization DOF with P-QND, and they can also divide
the states of the four-photon systems into two groups in the spatial-mode DOF with
S-QND \cite{HEPPECP}.

(1) After the X-quadrature measurement is performed on the coherent beam, the
results of the P-QND and S-QND showed that the polarization DOF of the
photon pair $AC$ is in an odd-parity mode and the spatial-mode DOF
of the photon pair $BD$ is also in an odd-parity mode. The state of the
four-photon system $ABCD$ is projected to $|\Phi_1\rangle$ with the
probability of $p(1)=4|\alpha\beta\gamma\delta|^2$. Here
\begin{eqnarray}                           
|\Phi_1\rangle \!&=&\!
\frac{1}{2}(|HHVV\rangle+|VVHH\rangle)_{ABCD}\nonumber\\
 &&\! \otimes (|a_2b_2c_1d_1\rangle+|a_1b_1c_2d_2\rangle).\label{eq10}
\end{eqnarray}
Then, Alice and Bob perform the Hadamard operations on the
polarization and spatial-mode DOFs of the two photons $C$ and $D$,
respectively. If the detection shows that the polarization and
spatial-mode DOFs of the two photons $C$ and $D$ are both in the
even-parity mode, the photon pair $AB$ in the maximally
hyperentangled Bell state
$|\phi\rangle_{AB}=\frac{1}{2}(|HH\rangle+|VV\rangle)_{AB}
(|a_1b_1\rangle+|a_2b_2\rangle)$ is obtained. If the detection shows
that the spatial-mode (polarization) DOF of the two photons $C$ and
$D$ is in an odd-parity mode, Bob performs a local phase-flip
operation $\sigma^S_z$ ($\sigma^P_z$) on photon $B$ to obtain the
state $|\phi\rangle_{AB}$.

(2) After the X-quadrature measurement is performed on the coherent beam, the
results of the P-QND and S-QND showed that the polarization DOF of the
photon pair $AC$ is in an even-parity mode and the spatial-mode DOF
of the photon pair $BD$ is also in an even-parity mode. The state of the
four-photon system $ABCD$ is projected to $|\Phi_2\rangle$
with the probability of
$p'(1)_1=(|\alpha|^4+|\beta|^4)(|\gamma|^4+|\delta|^4)$. Here
\begin{eqnarray}                           
|\Phi_2\rangle\!&=&\!\frac{1}{\sqrt{p'(1)_1}}(\alpha^2|HHHH\rangle
+\beta^2|VVVV\rangle)_{ABCD}\nonumber\\
&&\!\otimes
(\gamma^2|a_1b_1c_1d_1\rangle+\delta^2|a_2b_2c_2d_2\rangle).\label{eq12}
\end{eqnarray}
After Alice and Bob perform the Hadamard operations and detections
on the polarization and spatial-mode DOFs of the two photons $C$ and $D$
and the conditional local phase-flip operations $\sigma^S_z$
($\sigma^P_z$) on photon $B$, the state of the two-photon system $AB$ is
projected to $|\phi_2\rangle_{AB}$. Here
\begin{eqnarray}                           
|\phi_2\rangle_{AB}\!&=&\!\!\frac{1}{\sqrt{p'(1)_1}}(\alpha^2|HH\rangle+\beta^2|VV\rangle)_{AB}\nonumber\\
&&\! \otimes
(\gamma^2|a_1b_1\rangle+\delta^2|a_2b_2\rangle).\label{eq12}
\end{eqnarray}
This is a partially hyperentangled Bell state with less entanglement, and it can
be distilled to the maximally hyperentangled Bell state with another
round of the hyper-ECP process.

(3) After the X-quadrature measurement is performed on the coherent beam, the
results of the P-QND and  S-QND showed that the polarization DOF of the
photon pair $AC$ is in an odd-parity mode and the spatial-mode DOF
of the photon pair $BD$ is in an even-parity mode. The state of the
four-photon system $ABCD$ is projected to $|\Phi_3\rangle$ with the
probability of $p'(1)_2=2|\alpha\beta|^2(|\gamma|^4+|\delta|^4)$.
Here
\begin{eqnarray}                           
|\Phi_3\rangle \! &=& \! \frac{1}{\sqrt{2(|\gamma|^4+|\delta|^4)}}(|HHVV\rangle+|VVHH\rangle)_{ABCD}\nonumber\\
&&\!
\otimes(\gamma^2|a_1b_1c_1d_1\rangle+\delta^2|a_2b_2c_2d_2\rangle),
\end{eqnarray}
After Alice and Bob perform the Hadamard operations and detections
on the polarization and spatial-mode DOFs of the two photons $C$ and $D$
and the conditional local phase-flip operations $\sigma^S_z$
($\sigma^P_z$) on photon $B$, the state of the two-photon system $AB$ is
projected to $|\phi_3\rangle_{AB}$. Here
\begin{eqnarray}                           
|\phi_3\rangle_{AB}\!&=&\!\frac{1}{\sqrt{2(|\gamma|^4+|\delta|^4)}}(|HH\rangle+|VV\rangle)_{AB}\nonumber\\
&&\! \otimes(\gamma^2|a_1b_1\rangle+\delta^2|a_2b_2\rangle).
\end{eqnarray}
This is a partially hyperentangled Bell state with the polarization DOF in a maximally
entangled Bell state, and it can be distilled to the maximally hyperentangled Bell state
with another round of the hyper-ECP process.

(4) After the X-quadrature measurement is performed on the coherent beam, the
results of the P-QND and S-QND showed that the polarization DOF of the
photon pair $AC$ is in an even-parity mode and the spatial-mode DOF
of the photon pair $BD$ is in an odd-parity mode. The state of the
four-photon system $ABCD$ is projected to $|\Phi_4\rangle$ with the
probability of $p'(1)_3=2|\gamma\delta|^2(|\alpha|^4+|\beta|^4)$.
Here
\begin{eqnarray}                           
|\Phi_4\rangle\!\!&=&\!\!\frac{1}{\sqrt{2(|\alpha|^4\!+\!|\beta|^4)}}
(\alpha^2|HHHH\rangle\!+\!\beta^2|VVVV\rangle)_{\!ABCD}\nonumber\\
&& \otimes(|a_1b_1c_2d_2\rangle+|a_2b_2c_1d_1\rangle).
\end{eqnarray}
After Alice and Bob perform the Hadamard operations and detections
on the polarization and spatial-mode DOFs of the two photons $C$ and
$D$ and the conditional local phase-flip operations $\sigma^S_z$
($\sigma^P_z$) on photon $B$, the state of the two-photon system
$AB$ is projected to $|\phi_4\rangle_{AB}$. Here
\begin{eqnarray}                           
|\phi_4\rangle_{AB}\!\!&=&\!\!\frac{1}{\sqrt{2(|\alpha|^4+|\beta|^4)}}(\alpha^2|HH\rangle+\beta^2|VV\rangle)_{AB}\nonumber\\
&&\!\! \otimes(|a_1b_1\rangle+|a_2b_2\rangle).
\end{eqnarray}
This is a partially hyperentangled Bell state with the spatial-mode DOF in a maximally
entangled Bell state, and it can be distilled to the maximally hyperentangled Bell state
with another round of the hyper-ECP process.

\emph{The second round of the hyper-ECP} --- In the cases (2)$-$(4) of
the first round, the two-photon system $AB$ is projected to a
partially hyperentangled Bell state, which requires the second round
of the hyper-ECP process.

In the second round of the hyper-ECP process, two identical photon
pairs $AB$ and $A'B'$ are required. Alice and Bob perform the same
operations on their photon pairs $AA'$ and $BB'$ as they did in the
first round of the hyper-ECP.

(1') In the case (2) of the first round, the photon pairs $AB$ and
$A'B'$ are in the states $|\phi_2\rangle_{AB}$ and
$|\phi_2\rangle_{A'B'}$. After the P-QND and S-QND are performed on
the two-photon systems $AA'$ and $BB'$, respectively, the four cases
in the first round are all obtained. Therefore, after Alice and Bob
perform the Hadamard operations and detections on the polarization
and spatial-mode DOFs of the two photons $A'$ and $B'$ and the conditional
local phase-flip operations $\sigma^S_z$ ($\sigma^P_z$) on photon
$B$, the maximally hyperentangled Bell state $|\phi\rangle_{AB}$ is
obtained with the probability of
$p(2)_1=4|\alpha\beta\gamma\delta|^4/[(|\alpha|^4+|\beta|^4)(|\gamma|^4+|\delta|^4)]$.
The other three partially hyperentangled Bell states
$|\phi_2\rangle^1_{AB}$, $|\phi_2\rangle^2_{AB}$, and
$|\phi_2\rangle^3_{AB}$ can be obtained with the probabilities of
$p'(2)^1_1=(|\alpha|^8+|\beta|^8)(|\gamma|^8+|\delta|^8)/[(|\alpha|^4+|\beta|^4)(|\gamma|^4+|\delta|^4)]$,
$p'(2)^2_1=2|\alpha\beta|^4(|\gamma|^8+|\delta|^8)/[(|\alpha|^4+|\beta|^4)(|\gamma|^4+|\delta|^4)]$,
and
$p'(2)^3_1=2|\gamma\delta|^4(|\alpha|^8+|\beta|^8)/[(|\alpha|^4+|\beta|^4)(|\gamma|^4+|\delta|^4)]$,
respectively. Here
\begin{eqnarray}                           
|\phi_2\rangle^1_{AB}\!\!&=&\!\!\frac{1}{\sqrt{(|\alpha|^8+|\beta|^8)(|\gamma|^8+|\delta|^8)}}(\alpha^4|HH\rangle\nonumber\\
&&\!\! +\beta^4|VV\rangle)_{AB}\otimes
(\gamma^4|a_1b_1\rangle+\delta^4|a_2b_2\rangle),\nonumber\\
|\phi_2\rangle^2_{AB}\!\!&=&\!\!\frac{1}{\sqrt{2(|\gamma|^8+|\delta|^8)}}(|HH\rangle+|VV\rangle)_{AB}\nonumber\\
&&\!\!
\otimes(\gamma^4|a_1b_1\rangle+\delta^4|a_2b_2\rangle),\nonumber\\
|\phi_2\rangle^3_{AB}\!\!&=&\!\!\frac{1}{\sqrt{2(|\alpha|^8+|\beta|^8)}}(\alpha^4|HH\rangle+\beta^4|VV\rangle)_{AB}\nonumber\\
&&\!\! \otimes(|a_1b_1\rangle+|a_2b_2\rangle). \label{eq12}
\end{eqnarray}
If the photon pair $AB$ is projected to a partially hyperentangled Bell state, another
round of the hyper-ECP process is required.

(2') In the case (3) of the first round, the photon pairs $AB$ and
$A'B'$ are in the states $|\phi_3\rangle_{AB}$ and
$|\phi_3\rangle_{A'B'}$. After the P-QND and S-QND are performed on
the two-photon systems $AA'$ and $BB'$, respectively, Alice and Bob
pick up the case that the result of S-QND is in an odd-parity mode (the
result of P-QND is in either an even-parity mode or an odd-parity
mode). After Alice and Bob perform the Hadamard operations and
detections on the polarization and spatial-mode DOFs of the two photons
$A'$ and $B'$ and the conditional local phase-flip operations $\sigma^S_z$
($\sigma^P_z$) on photon $B$, the maximally hyperentangled Bell
state $|\phi\rangle_{AB}$ is obtained with the probability of
$p(2)_2=4|\gamma\delta|^4|\alpha\beta|^2/(|\gamma|^4+|\delta|^4)$.
If the result of S-QND is in an even-parity mode (the result of
P-QND is in either an even-parity mode or an odd-parity mode), the
partially hyperentangled Bell state $|\phi_3\rangle^1_{AB}$ is
obtained with the probability of
$p'(2)_2=2(|\gamma|^8+|\delta|^8)|\alpha\beta|^2/(|\gamma|^4+|\delta|^4)$.
Here
\begin{eqnarray}                           
|\phi_3\rangle^1_{AB}&=&\frac{1}{\sqrt{2(|\gamma|^8+|\delta|^8)}}(|HH\rangle+|VV\rangle)_{AB}\nonumber\\
&& \otimes(\gamma^4|a_1b_1\rangle+\delta^4|a_2b_2\rangle).
\end{eqnarray}
If the photon pair $AB$ is projected to a partially hyperentangled Bell state, another
round of the hyper-ECP process is required.

(3') In the case (4) of the first round, the photon pairs $AB$ and
$A'B'$ are in the states $|\phi_4\rangle_{AB}$ and
$|\phi_4\rangle_{A'B'}$. After the P-QND and S-QND are performed on
the two-photon systems $AA'$ and $BB'$, respectively, Alice and Bob
pick up the case that the result of P-QND is in an odd-parity mode (the
result of S-QND is in either an even-parity mode or an odd-parity
mode). After Alice and Bob perform the Hadamard operations and
detections on the polarization and spatial-mode DOFs of the two photons
$A'$ and $B'$ and the conditional local phase-flip operations $\sigma^S_z$
($\sigma^P_z$) on photon $B$, the maximally hyperentangled Bell
state $|\phi\rangle_{AB}$ is obtained with the probability of
$p(2)_3=4|\alpha\beta|^4|\gamma\delta|^2/(|\alpha|^4+|\beta|^4)$. If
the result of P-QND is in an even-parity mode (the result of S-QND
is in either an even-parity mode or an odd-parity mode), the
partially hyperentangled Bell state $|\phi_4\rangle^1_{AB}$ is
obtained with the probability of
$p'(2)_3=2(|\alpha|^8+|\beta|^8)|\gamma\delta|^2/(|\alpha|^4+|\beta|^4)$.
Here
\begin{eqnarray}                           
|\phi_4\rangle^1_{AB}&=&\frac{1}{\sqrt{2(|\alpha|^8+|\beta|^8)}}(\alpha^4|HH\rangle+\beta^4|VV\rangle)_{AB}\nonumber\\
&& \otimes(|a_1b_1\rangle+|a_2b_2\rangle).
\end{eqnarray}
If the photon pair $AB$ is projected to a partially hyperentangled Bell state, another
round of the hyper-ECP process is required.

\begin{center}                          
\begin{figure}[!h]
\includegraphics*[width=3in]{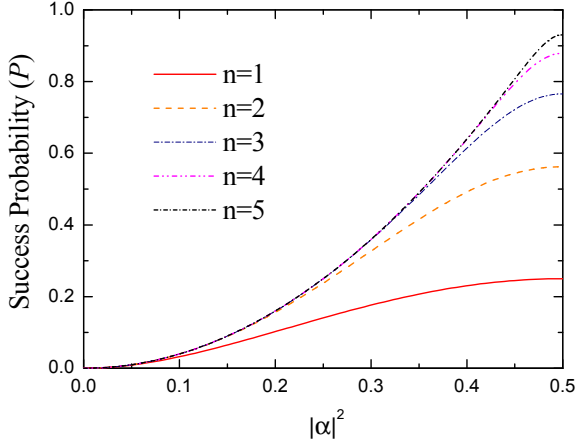}
\caption{Success probability of the hyper-ECP for a pair of
partially hyperentangled Bell states under the iteration numbers
($n$) \cite{HEPPECP,HECPLixhOE}. The parameters of the
polarization-spatial partially hyperentangled Bell state are chosen
as $|\alpha|=|\gamma|$ and $|\beta|=|\delta|$. }\label{figure3.5}
\end{figure}
\end{center}

\emph{Success probability} --- The success probabilities of each
round of the hyper-ECP process are $p(1)$,
$p(2)=p(2)_1+p(2)_2+p(2)_3$, \ldots, respectively. After $n$ rounds
of the hyper-ECP process, the total success probability of the
hyper-ECP is $P=\sum_np(n)$. The relation between the success
probability $P$ and the parameter $|\alpha|^2$ under the iteration
numbers $n$ is shown in Fig.~\ref{figure3.5} for the case with
$|\alpha|<|\beta|$ ($|\alpha|=|\gamma|$ and $|\beta|=|\delta|$). It
shows that the success probability of the hyper-ECP is improved
largely by iterative application of the hyper-ECP process several
times.

Now, we can see that the success probability of the hyper-ECP is
improved by using the parity-check QNDs, resorting to nonlinear
optical elements. The P-QND and S-QND in this protocol can also be
constructed with other nonlinear optical elements with similar
effect, such as cavity-NV-center system \cite{HEPPECP}, quantum-dot-cavity
system \cite{HECP2}, and so on. In fact, this hyper-ECP
is implemented by concentrating  the polarization and spatial-mode
DOFs independently. If the swap gate is introduced, the success
probability of each round of the hyper-ECP will be greatly improved
by transferring the useful information between the nonlocal
partially hyperentangled Bell states in cases (3) and (4) in the
first round of the hyper-ECP. The detail of the highly efficient
two-step hyper-ECP with quantum swap gates is introduced in
Ref.~\cite{HECP3}.

\subsection{Hyper-ECP for polarization-time-bin hyperentangled Bell state}
\label{sec3.2.2}

The time-bin DOF is a simple and conventional DOF, and it is also
very useful in quantum information processing. Here, we introduce a
hyper-ECP for unknown polarization-time-bin hyperentangled Bell
state with the Schmidt projection method \cite{HECP4}.

Two nonlocal photon pairs $AB$ and $CD$ are required in this
proposal, and they are in the partially hyperentangled states
$|\phi\rangle_{AB}$ and $|\phi\rangle_{CD}$, respectively. Here,
\begin{eqnarray}                           
|\phi\rangle_{AB}&=&(\alpha|HH\rangle+\beta|VV\rangle)\otimes(\delta|SS\rangle+\eta|LL\rangle)_{AB},\nonumber\\
|\phi\rangle_{CD}&=&(\alpha|HH\rangle+\beta|VV\rangle)\otimes(\delta|SS\rangle+\eta|LL\rangle)_{CD}.\nonumber\\
\end{eqnarray}
Here $|S\rangle$ and $|L\rangle$ represent two time-bins early and
late, respectively, and there is a time interval $\Delta t$ between
the two time-bins. The photons $A$ and $C$ belong to Alice, and the
photons $B$ and $D$ belong to Bob. $\alpha$, $\beta$, $\delta$, and
$\eta$ are four unknown parameters, and they satisfy the
normalization condition
$|\alpha|^2+|\beta|^2=|\delta|^2+|\eta|^2=1$.

\begin{center}                         
\begin{figure}[!h]
\includegraphics*[width=8.5cm]{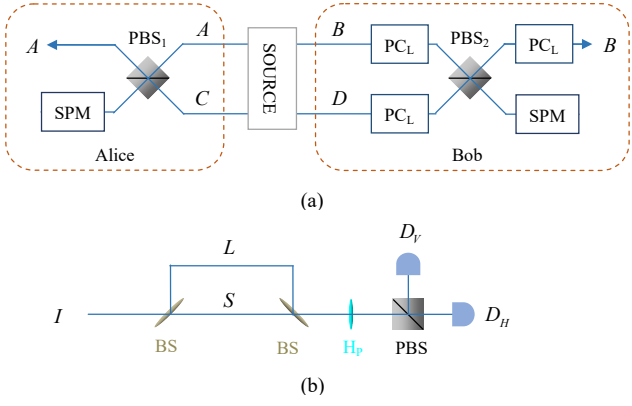}
\caption{(a) Schematic diagram of the polarization-time-bin
hyper-ECP for partially hyperentangled Bell states with unknown
parameters  \cite{HECP4}, resorting to the Schmidt projection
method. PC$_L$ (PC$_S$) represents a Pockels cell, which is used to
perform polarization bit-flip operation on the $L$ ($S$) component.
(b) Schematic diagram of a single-photon measurement
(SPM).}\label{figure3}
\end{figure}
\end{center}

The initial state of the  four-photon system $ABCD$ is
$|\Phi\rangle_0=|\phi\rangle_{AB}\otimes|\phi\rangle_{CD}$. First,
Alice and Bob perform the polarization bit-flip operations
($\sigma^P_x$) with half-wave plates (X) and the time-bin bit-flip
operations ($\sigma^T_x=|S\rangle\langle L|+|L\rangle\langle S|$)
with the active switches on the two photons $C$ and $D$. The state
of the four-photon system $ABCD$ is transformed into
$|\Phi\rangle_1$. Here
\begin{eqnarray}                           
|\Phi\rangle_1&=&[(\alpha^2|HHVV\rangle+\beta^2|VVHH\rangle)\nonumber\\
&&+\alpha\beta(|HHHH\rangle+|VVVV\rangle)]\nonumber\\
&&\otimes[(\delta^2|SSLL\rangle+\eta^2|LLSS\rangle)\nonumber\\
&&+\delta\eta(|SSSS\rangle+|LLLL\rangle)]_{ABCD}.
\end{eqnarray}

Subsequently, Alice and Bob put the photon pairs $AC$ and $BD$ into
the quantum circuit shown in Fig.~\ref{figure3}a. That is, Alice
put the two photons $A$ and $C$ into the polarization beam splitter
PBS$_1$, which can perform a parity-check measurement on the
polarization DOF of the photon pair $AC$. Alice picks up the
even-parity polarization state with only one photon exiting from
each output port of PBS$_1$, and the state of the four-photon system
will be transformed from $|\Phi\rangle_1$ to $|\Phi\rangle_2$ with
the probability of $2|\alpha\beta|^2$. Here
\begin{eqnarray}                           
|\Phi\rangle_2\!\!&=&\!\!\frac{1}{\sqrt{2}}(|HHHH\rangle+|VVVV\rangle)\nonumber\\
\!\!&&\otimes[(\delta^2|SSLL\rangle+\eta^2|LLSS\rangle)\nonumber\\
\!\!&&+\delta\eta(|SSSS\rangle+|LLLL\rangle)]_{ABCD}.
\end{eqnarray}

At the same time, Bob puts the two photons $B$ and $D$ into two
Pockels cells (PCs), which can perform the polarization bit-flip
operations on the photon system at a specific time. That is, PC$_L$
(PC$_S$) only performs the polarization bit-flip operation on the
$L$ ($S$) component. Then the state of the four-photon system is
changed from $|\Phi\rangle_2$ to $|\Phi\rangle_3$. Here
\begin{eqnarray}                           
|\Phi\rangle_3\!\!&=&\!\!\frac{1}{\sqrt{2}}[\delta^2(|H^SH^SH^LV^L\rangle+|V^SV^SV^LH^L\rangle)\nonumber\\
\!\!&&+\eta^2(|H^LV^LH^SH^S\rangle+|V^LH^LV^SV^S\rangle)\nonumber\\
\!\!&&+\delta\eta(|H^SH^SH^SH^S\rangle+|V^SV^SV^SV^S\rangle\nonumber\\
\!\!&&+|H^LV^LH^LV^L\rangle+|V^LH^LV^LH^L\rangle)]_{ABCD}.\nonumber\\
\end{eqnarray}
The superscript $L$ ($S$) represents the time-bin late (early). Then
Bob can perform the time-bin parity-check measurement on the two
photons $B$ and $D$ with PBS$_2$, and he also picks up the
even-parity polarization state with only one photon exiting from
each output port of PBS$_2$. After another PC$_L$ is performed on
photon $B$, the four-photon system will be projected into the state
$|\Phi\rangle_4$ with the probability of
$4|\alpha\beta\delta\eta|^2$. Here
\begin{eqnarray}                           
|\Phi\rangle_4\!\!&=&\!\!\frac{1}{2}(|H^SH^SH^SH^S\rangle+|V^SV^SV^SV^S\rangle\nonumber\\
\!\!&&+|H^LH^LH^LV^L\rangle+|V^LV^LV^LH^L\rangle)_{ABCD}.\nonumber\\
\end{eqnarray}

The last step of this hyper-ECP is to detect the two photons $C$ and
$D$ with the single-photon measurement (SPM) shown in
Fig.~\ref{figure3}b, which can transform the state
$|\Phi\rangle_4$ to one of the maximally hyperentangled Bell states
$|\psi^{\pm\pm}\rangle_{AB}$. Here
\begin{eqnarray}                           
|\psi^{\pm\pm}\rangle_{AB} = \frac{1}{2}(|HH\rangle\pm|VV\rangle)_{AB}\otimes(|SS\rangle\pm|LL\rangle)_{AB}.\nonumber\\
\end{eqnarray}

In the quantum circuit of SPM shown in Fig.~\ref{figure3}b,
the length difference between the two arms of beam splitters (BS$_1$ and BS$_2$)
is $c\Delta t$, which is used as an unbalanced interferometer (UI).
Here $c$ is the speed of the photons. The effect of this UI is
\begin{eqnarray}                           
|X^L\rangle&\rightarrow&\frac{1}{\sqrt{2}}(|X^{LS}\rangle+|X^{LL}\rangle),\nonumber\\
|X^S\rangle&\rightarrow&\frac{1}{\sqrt{2}}(|X^{SS}\rangle+|X^{SL}\rangle).
\end{eqnarray}
Here, $X$ represents $H$ or $V$, and $X^{ij}$ ($i,j=L,S$) represents
the time-bin state $X^i$ passing through the arm $j$ of the UI. After
the two photons $C$ and $D$ pass through UI, the state of the
four-photon system is changed to
\begin{eqnarray}                           
&&|H^SH^S\rangle_{AB}\!\otimes\!(|H^{SS}\rangle\!+\!|H^{SL}\rangle)_C\!\otimes\!(|H^{SS}\rangle\!+\!|H^{SL}\rangle)_D\nonumber\\
&&+|V^SV^S\rangle_{AB}\!\otimes\!(|V^{SS}\rangle\!+\!|V^{SL}\rangle)_C\!\otimes\!(|V^{SS}\rangle\!+\!|V^{SL}\rangle)_D\nonumber\\
&&+|H^LH^L\rangle_{AB}\!\otimes\!(|H^{LS}\rangle\!+\!|H^{LL}\rangle)_C\!\otimes\!(|H^{LS}\rangle\!+\!|H^{LL}\rangle)_D\nonumber\\
&&+|V^LV^L\rangle_{AB}\!\otimes\!(|V^{LS}\rangle\!+\!|V^{LL}\rangle)_C\!\otimes\!(|V^{LS}\rangle\!+\!|V^{LL}\rangle)_D.\nonumber\\
\end{eqnarray}
The time-bin components $LS$ and $SL$ will arrive at the same time,
and the time-bin components $LL$ and $SS$ will arrive at a later
time and an earlier time, respectively. In order to obtain the
maximally hyperentangled Bell state, Alice and Bob only pick up the
states arriving at the middle time $LS$ and $SL$.  After the two
photons $C$ and $D$ are detected, the local polarization
operation and time-bin operation have to be performed
on the photon $B$ to obtain the maximally hyperentangled  Bell state
$|\psi^{++}\rangle_{AB}$, which is shown in Table \ref{table1}.

\begin{table}[htb]
\centering
\caption{The relation between measurement results of $CD$  in
the middle time slot, the final states of $AB$, and the local operations to obtain the maximally hyperentangled Bell state
$\vert \psi^{++}\rangle_{AB}$.}
\begin{tabular}{ccc}
\hline
    Detection ($CD$)             &          State of $AB$      &        Local operation           \\
   \hline

  $HH$   &        $|\psi^{++}\rangle^{AB}$        &       $I$             \\

  $HV$    &       $|\psi^{--}\rangle^{AB}$        &       $\sigma^T_z$, $\sigma^P_z$              \\

  $VH$    &       $|\psi^{-+}\rangle^{AB}$           &    $\sigma^P_z$               \\

  $VV$    &       $|\psi^{+-}\rangle^{AB}$          &     $\sigma^T_z$        \\ \hline
\end{tabular}\label{table1}
\end{table}

The success probability of this hyper-ECP is
$P=|\alpha\beta\gamma\delta|^2$, which is a quarter of the one for
the polarization-spatial hyperentangled Bell sate. An improved SPM
is also introduced by Li and Ghose \cite{HECP4}. With the improved
SPM, the success probability of this hyper-ECP can be enhanced to
$P=4|\alpha\beta\gamma\delta|^2$.

\section{Hyperentanglement  purification}

\subsection{History of entanglement purification}

Entanglement purification is an important passive way for depressing
the influence of noise on nonlocal quantum systems and it is an
indispensable technique in quantum repeaters. Generally speaking, it
is used to distill some nonlocal entangled systems in a
high-fidelity entangled state from a set of nonlocal entangled
systems in a mixed state with less entanglement \cite{EPP1}. In
1996, Bennett et al. \cite{EPP1} presented the first entanglement
purification protocol (EPP) which is used to purify a Werner state
based on quantum controlled-not (CNOT) gates. Subsequently, Deutsch
et al. \cite{EPPDeutsch} improved this EPP by adding  two special
unitary operations for each particle. In 2001, Pan, Simon, and
Zellinger \cite{EPP2} proposed an EPP with linear-optics elements
for ideal entangled photon sources. In 2002, Simon and Pan
\cite{EPP3} proposed an EPP for two entangled photons from a SPDC
source with two PBSs, and Pan et al. \cite{EPP4} demonstrated this
EPP in experiment in 2003. Based on the cross-Kerr nonlinearities,
an efficient polarization EPP was proposed by Sheng et al.
\cite{EPPsheng1} in 2008. In this scheme, the parties in quantum
communication can increase the entanglement and improve the fidelity
of quantum states by repeatedly performing the purification
protocols. In 2011, Wang et al. \cite{EPPWangCQIP} proposed an
interesting EPP using cross-Kerr nonlinearity by identifying the
intensity of probe coherent beams. Also, Wang et al.
\cite{EPPWangCPRA} presented an EPP for electron-spin entangled
states using quantum-dot spin and microcavity coupled systems.  In
2013, Sheng et al. \cite{EPPhybridSheng} presented an
hybrid EPP for quantum repeaters. In 2016, Zhou and Sheng
\cite{analysissheng7} proposed an EPP for logic-qubit entanglement.


Entanglement purification makes great progress since the concept of
deterministic entanglement purification was introduced originally by
Sheng and Deng \cite{EPPsheng2} in 2010. In this year, they
\cite{EPPsheng2} presented a two-step deterministic EPP for
polarization entanglement with the hyperentanglement in both the
spatial-mode and the frequency DOFs of photon pairs. Subsequently,
Sheng and Deng \cite{EPPsheng3} and Li \cite{EPPlixh} independently
proposed the one-step deterministic EPP  for polarization
entanglement with only the spatial entanglement of photon pairs,
resorting to linear-optical elements only.  In 2011, Deng
\cite{EPPdeng} extended the deterministic entanglement purification
to multipartite entanglement with the spatial entanglement or the
frequency entanglement of photon systems. Moreover, he
\cite{EPPdeng} showed that a  deterministic EPP does not require the
photon systems entangled in the polarization DOF, but one error-free
DOF. These one-step deterministic EPPs
\cite{EPPsheng3,EPPlixh,EPPdeng} can purify the polarization
entanglement with one step, resorting to linear-optical elements,
and the polarization errors are totally converted into the ambiguity
of spatial modes when the two photons in each pair are originally
entangled in spatial DOF which has been exploited to produce  a (100
$\times$ 100)-dimensional entanglement \cite{PNAS}. In 2014, Sheng
and Zhou \cite{DEPPShengLPL} also described another good
deterministic EPP for polarization entanglement with time-bin
entanglement. In 2015, Sheng and Zhou \cite{DEPPSR} proposed the
deterministic entanglement distillation for secure double-server
blind quantum computation.  The deterministic EPPs
\cite{EPPsheng2,EPPsheng3,EPPlixh,EPPdeng} are far different from
the conventional EPPs
\cite{EPP1,EPPDeutsch,EPP2,EPP3,EPP4,EPPsheng1,EPPWangCQIP,EPPWangCPRA,EPPhybridSheng}
as they work in a completely deterministic way, not in a
probabilistic way, and they can reduce the quantum resource
sacrificed largely. They are very useful in quantum repeaters.

The purification of nonlocal hyperentangled quantum systems is more
complex than that of entangled systems in one DOF. In 2013, Ren and
Deng \cite{HEPPECP} presented the first hyperentanglement
purification protocol (hyper-EPP) for two-photon systems in
polarization-spatial hyperentangled states, and it is very useful in
the high-capacity quantum repeaters with hyperentanglement. In 2014,
Ren et al.  \cite{HEPP1} gave the two-step hyper-EPP for
polarization-spatial hyperentangled states with the
quantum-state-joining method \cite{QSJM}, and it has a far higher
efficiency. In 2016,  Wang et al. \cite{HEPP2wang} presented
the first hyper-EPP for two-photon six-qubit hyperentangled systems in
three DOFs, and they  showed that using SWAP gates is a universal
method for hyper-EPP for nonlocal hyperentangled quantum states in
both the polarization DOF and multiple-longitudinal-momentum DOFs to
obtain a high yield (efficiency), as well as other hyperentanglement
with more than two DOFs. In 2015, Wang et al.
\cite{HEPPadd,HEPPaddWW} proposed two novel hyper-EPPs by combining
both the conventional EPP and photon-loss amplification on
hyperentangled photon systems assisted by local entanglement
resource.


\subsection{Two-step hyper-EPP}

Here, we introduce the principle of the two-step hyper-EPP
\cite{HEPP1} for polarization-spatial hyperentangled Bell state with
the quantum-state-joining method (QSJM).

\begin{center}                       
\begin{figure}[!h]
\includegraphics*[width=2.5in]{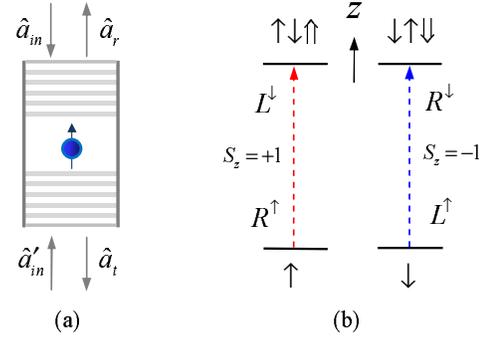}
\caption{(a) A double-sided QD-cavity system. (b) The spin-dependent
optical transitions of a negatively charged exciton $X^-$ with
circularly polarized photons \cite{HEPP1}. $L^\uparrow$
($L^\downarrow$) and $R^\uparrow$ ($R^\downarrow$) represent the
left and the right circularly polarized lights with their input
directions parallel (antiparallel) with z direction, respectively.
}\label{figure3.8}
\end{figure}
\end{center}

For simplification, we only discuss the principle of the  two-step
hyper-EPP  for mixed hyperentangled Bell states with polarization
bit-flip errors and spatial-mode phase-flip errors \cite{HEPP1},
resorting to the polarization-spatial phase-check QND (P-S-QND) and
QSJM. This hyper-EPP can be used for nonlocally mixed hyperentangled
Bell states with arbitrary errors in the two DOFs. In this
hyper-EPP, the P-S-QND is used to distinguish the Bell state with a
relative phase $0$ from the one with a relative phase $\pi$ in both
the polarization and the spatial-mode DOFs. The QSJM is used to
combine the polarization state of photon $A$ and the spatial-mode
state of photon $B$ into an output single-photon state. Both these
two basic gate elements are constructed with the nonlinearity of a
quantum dot (QD) embedded in a double-sided optical microcavity
(double-sided QD-cavity system, shown in Fig.~\ref{figure3.8}a).

\emph{The input-output optical property of the double-sided
QD-cavity system}
--- The two distributed Bragg reflectors of the double-sided optical
microcavity are partially reflective, and they are low loss for
on-resonance transmission in both the two polarization modes. When
an excess electron is injected into a QD \cite{QD5}, the negatively
charged exciton $X^-$, which consists of two antiparallel electrons
bound to one hole, will be created by resonantly absorbing a
circularly polarized light, according to the spin-dependent
transition rule with Pauli's exclusion principle \cite{QD6,QD1}
(shown in Fig.~\ref{figure3.8}b). That is, a circularly polarized
photon with the spin $S_z=+1$ is absorbed to create the negatively
charged exciton $X^-$ in the state
$|\uparrow\downarrow\Uparrow\rangle$ for the excess electron spin
$|\uparrow\rangle$, and a circularly polarized photon with the spin
$S_z=-1$ is absorbed to create the negatively charged exciton $X^-$
in the state $|\downarrow\uparrow\Downarrow\rangle$ for the excess
electron spin $|\downarrow\rangle$. The state $|\Uparrow\rangle$
($|\Downarrow\rangle$) represents the heavy-hole spin
$|+\frac{3}{2}\rangle$ ($|-\frac{3}{2}\rangle$), and the state
$|\uparrow\rangle$ ($|\downarrow\rangle$) represents the electron
spin $|+\frac{1}{2}\rangle$ ($|-\frac{1}{2}\rangle$).

The input-output optical property of the double-sided QD-cavity
system can be described by the Heisenberg equations of motion for
the cavity field operator $\hat{a}$ and $X^-$ dipole operator
$\hat{\sigma}_-$ in the interaction picture \cite{QD7},
\begin{eqnarray}                           
\frac{{\rm d}\hat{a}}{{\rm d}t}\!\!&=&\!\!-[{\rm i}(\omega_{\rm c}-\omega)+\kappa+\!\frac{\kappa_{\rm s}}{2}]\hat{a}-g\hat{\sigma}_-
 \!-\!\sqrt{\kappa}\,\hat{a}_{\rm in}\!-\!\sqrt{\kappa}\,\hat{a}'_{\rm in}, \nonumber\\
\frac{{\rm d}\hat{\sigma}_-}{{\rm d}t}\!\!&=&\!\!-[{\rm i}(\omega_{X^-}-\omega)+\frac{\gamma}{2}]\hat{\sigma}_--g\hat{\sigma}_z\hat{a},
\end{eqnarray}
where $\omega$, $\omega_{\rm c}$, and $\omega_{X^-}$ are the frequencies
of the input photon, the cavity field mode, and the $X^-$
transition, respectively. $\kappa$ and $\kappa_{\rm s}/2$ are the decay
rates of the cavity field mode to the output photon and the side
leakage mode, respectively. $g$ is the coupling strength of the
negatively charged exciton $X^-$ and the cavity field mode.
$\gamma/2$ is the decay rate of negatively charged exciton $X^-$.
$\hat{a}_{\rm in}$ and $\hat{a}'_{\rm in}$ are the input field operators of
the double-sided QD-cavity system. $\hat{a}_{r}$ and $\hat{a}_{t}$
are the output field operators of the double-sided QD-cavity system.
These field operators satisfy the boundary condition
$\hat{a}_{r}=\hat{a}_{\rm in}+\sqrt{\kappa}\,\hat{a}$ and
$\hat{a}_{t}=\hat{a}'_{\rm in}+\sqrt{\kappa}\,\hat{a}$. In the weak
excitation limit ($\langle\hat{\sigma}_z\rangle=-1$), the reflection
coefficient ($r(\omega)$) and transmission coefficient ($t(\omega)$)
of the double-sided QD-cavity system can be expressed as
\cite{QD3,aninputoutput}
\begin{eqnarray}                           
r(\omega)\!&=&\!1+t(\omega),\nonumber\\
t(\omega)\!&=&\!\frac{-\kappa[{\rm i}(\omega_{X^-}-\omega)+\frac{\gamma}{2}]}{[{\rm i}(\omega_{X^-}\!-\!\omega)
+\!\frac{\gamma}{2}][{\rm i}(\omega_{\rm c}-\omega)+\kappa+\frac{\kappa_{\rm s}}{2}]+g^2}.\;\;\;\;\;\;
\end{eqnarray}

In the resonant condition ($\omega_{\rm c}=\omega_{X^-}=\omega$) with
$\kappa_{\rm s}\rightarrow0$, the reflection and transmission coefficients
are $r_0\rightarrow0$ and $t_0\rightarrow-1$ for $g=0$, and they are
$r\rightarrow1$ and $t\rightarrow0$ for the strong coupling regime
$g>(\kappa,\gamma)$. As the photonic circular polarization is
dependent on the direction of propagation, the photon with spin
$S_z=+1$ corresponds to the state $|R^\uparrow\rangle$ or
$|L^\downarrow\rangle$, and the photon with spin $S_z=-1$
corresponds to the state $|R^\downarrow\rangle$ or
$|L^\uparrow\rangle$. Here $L^\uparrow$ ($R^\uparrow$) or
$L^\downarrow$ ($R^\downarrow$) represents the input direction of
the left (right) circularly polarized light which is parallel or
antiparallel to the z direction, respectively, as shown in
Fig.~\ref{figure3.8}b. In this condition, the reflection and
transmission rules of the photonic polarization states are described
as follows,
\begin{eqnarray}                           
&&\!\!\!\!|R^\uparrow, i_2, \uparrow\rangle \rightarrow
|L^\downarrow, i_2, \uparrow\rangle,\;\;\;\;\;\,
|L^\downarrow, i_1, \uparrow\rangle \rightarrow |R^\uparrow, i_1, \uparrow\rangle,\nonumber\\
&& \!\!\!\!|R^\uparrow, i_2, \downarrow\rangle \rightarrow
-|R^\uparrow, i_1\downarrow\rangle,\;\;\; |L^\downarrow,
i_1,\downarrow\rangle \rightarrow -|L^\downarrow, i_2,
\downarrow\rangle,\nonumber\\
 &&\!\!\!\!|R^\downarrow, i_1,
\uparrow\rangle \rightarrow -|R^\downarrow, i_2,
\uparrow\rangle,\;\;\,
|L^\uparrow, i_2, \uparrow\rangle \rightarrow -|L^\uparrow, i_1, \uparrow\rangle,\nonumber\\
&& \!\!\!\!|R^\downarrow, i_1, \downarrow\rangle \rightarrow
|L^\uparrow, i_1, \downarrow\rangle,\;\;\;\;\;\, |L^\uparrow, i_2,
\downarrow\rangle \rightarrow |R^\downarrow, i_2,
\downarrow\rangle.\;\;\;\;\;\;\;\;\label{eq4}
\end{eqnarray}

\begin{center}                                 
\begin{figure}[!h]
\includegraphics*[width=2.8in]{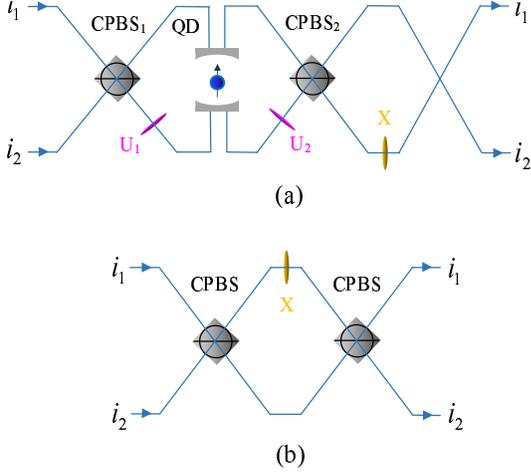}
\caption{(a) Schematic diagram of the quantum-state-joining method
(QSJM)  \cite{HEPP1}. CPBS represents a polarizing beam splitter in
the circular basis, which transmits the photon in the right-circular
polarization $|R\rangle$ and reflects the photon in the
left-circular polarization $|L\rangle$, respectively. U$_i$
($i=1,2$) represents a wave plate which is used to perform a
polarization phase-flip operation $U=-|R\rangle\langle
R|-|L\rangle\langle L|$ on a photon. X represents a half-wave plate
which is used to perform a polarization bit-flip operation
$\sigma^P_x=|R\rangle\langle L| + |L\rangle\langle R|$. (b)
Schematic diagram of the swap gate between the spatial-mode state
and the polarization state of a photon. }\label{figure3.9}
\end{figure}
\end{center}

\emph{Quantum-state-joining method} --- The QSJM is used to combine
the polarization state of photon $A$ and the spatial-mode state of
photon $B$ into an output single photon state. That is, the
polarization state of photon $A$ is transferred to the polarization
DOF of photon $B$. This QSJM is constructed with the nonlinearity of
a double-sided QD-cavity system \cite{HEPP1}, as shown in
Fig.~\ref{figure3.9}a. The initial state of the excess electron
spin in QD is
$\frac{1}{\sqrt{2}}(|\uparrow\rangle+|\downarrow\rangle)_e$. The
initial states of two photons $A$ and $B$ are $|\varphi_A\rangle =
(\alpha_1|R\rangle+\alpha_2|L\rangle)_A(\gamma_1|a_1\rangle+\gamma_2|a_2\rangle)$
and $|\varphi_B\rangle =
(\beta_1|R\rangle+\beta_2|L\rangle)_B(\delta_1|b_1\rangle+\delta_2|b_2\rangle)$,
respectively.

First, the two spatial modes of photon $A$ are put into CPBS$_1$, U$_1$, QD, U$_2$, CPBS$_2$, and X in sequence,
as shown in Fig.~\ref{figure3.9}a.
The state of the system $Ae$ is transformed from $|\varphi_{Ae}\rangle_0$ to
$|\varphi_{Ae}\rangle_1$.  Here
\begin{eqnarray}                           
|\varphi_{Ae}\rangle_0\!\!&=&\!\!|+\rangle_e\otimes|\varphi_A\rangle,\nonumber\\
|\varphi_{Ae}\rangle_1\!\!&=&\!\!\frac{1}{\sqrt{2}}\big[|R\rangle_A(\alpha_1|\uparrow\rangle+\alpha_2|\downarrow\rangle)_e
+|L\rangle_A(\alpha_2|\uparrow\rangle\nonumber\\
&&
+\alpha_1|\downarrow\rangle)_e\big](\gamma_1|a_1\rangle+\gamma_2|a_2\rangle).\label{eq5}
\end{eqnarray}
Then the polarization state of photon $A$ is measured in the
orthogonal basis $\{|R\rangle,|L\rangle\}$.
If the polarization state of photon $A$ is projected to
$|R\rangle$, the excess electron spin state in QD is
$|\phi\rangle_e=(\alpha_1|\uparrow\rangle+\alpha_2|\downarrow\rangle)_e$.
Otherwise, the excess electron spin state in QD is
$|\phi'\rangle_e=(\alpha_2|\uparrow\rangle+\alpha_1|\downarrow\rangle)_e$.

Subsequently, after a Hadamard operation is performed on the excess
electron spin $e$ in QD, the two spatial modes of photon $B$ are put
into the quantum circuit shown in Fig.~\ref{figure3.9}a. Here the
Hadamard operation on the excess electron spin is
$|\uparrow\rangle_e\rightarrow\frac{1}{\sqrt{2}}(|\uparrow\rangle+|\downarrow\rangle)_e,
|\downarrow\rangle_e\rightarrow\frac{1}{\sqrt{2}}(|\uparrow\rangle-|\downarrow\rangle)_e$.
Then the state of the system  $Be$ is transformed from
$|\varphi_{Be}\rangle_1$ to $|\varphi_{Be}\rangle_2$. Here
\begin{eqnarray}                           
|\varphi_{Be}\rangle_1\!\!&=&\!\!|\varphi_B\rangle\otimes|\phi_e\rangle,\nonumber\\
|\varphi_{Be}\rangle_2\!\!&=&\!\!\big[\alpha'_1|\uparrow\rangle_e(\beta_1|R\rangle+\beta_2|L\rangle)_B+\alpha'_2|\downarrow\rangle_e
(\beta_2|R\rangle\nonumber\\
&&+\beta_1|L\rangle)_B\big](\delta_1|b_1\rangle+\delta_2|b_2\rangle),\;\;\;\;\;\;\;\;
\end{eqnarray}
where $\alpha'_1=\frac{1}{\sqrt{2}}(\alpha_1+\alpha_2)$ and
$\alpha'_2=\frac{1}{\sqrt{2}}(\alpha_1-\alpha_2)$.

Next, the Hadamard operations are performed on the
polarization DOF of photon $B$ and the excess electron spin $e$, respectively.
Then the two spatial modes of photon $B$ are put into the quantum circuit shown in Fig.~\ref{figure3.9}a
again, and the state of the system $Be$ is changed to
$|\varphi_{Be}\rangle_3$. Here
\begin{eqnarray}                          
|\varphi_{Be}\rangle_3\!\!&=&\!\!\big[\alpha_1|R\rangle_B(\beta'_1|\uparrow\rangle+\beta'_2|\downarrow\rangle)_e+\alpha_2|L\rangle_B
(\beta'_2|\uparrow\rangle\nonumber\\
\!\!&&+\beta'_1|\downarrow\rangle)_e\big]\otimes(\delta_1|b_1\rangle+\delta_2|b_2\rangle),
\end{eqnarray}
where $\beta'_1=\frac{1}{\sqrt{2}}(\beta_1+\beta_2)$ and
$\beta'_2=\frac{1}{\sqrt{2}}(\beta_1-\beta_2)$.

Finally, a Hadamard operation is performed on the excess electron spin $e$ again.
The state of the system $Be$ is transferred from $|\varphi_{Be}\rangle_3$ to
$|\varphi_{Be}\rangle_4$. Here
\begin{eqnarray}                           
|\varphi_{Be}\rangle_4\!\!&=&\!\!\big[\beta_1|\uparrow\rangle_e(\alpha_1|R\rangle+\alpha_2|L\rangle)_B+\beta_2|\downarrow\rangle_e
(\alpha_1|R\rangle\nonumber\\
\!\!&&-\alpha_2|L\rangle)_B\big](\delta_1|b_1\rangle+\delta_2|b_2\rangle).
\end{eqnarray}
The excess electron spin $e$ is measured in the orthogonal basis
$\{|\!\!\uparrow\rangle,|\!\!\downarrow\rangle\}$. If the detection
shows that the polarization state of photon $A$ is $|L\rangle$, a
polarization bit-flip operation ($\sigma^P_x=|R\rangle\langle L| +
|L\rangle\langle R|$) on photon $B$ is required. If the detection
shows that the excess electron spin state is
$|\!\!\downarrow\rangle_e$, a polarization  phase-flip operation
($\sigma^P_z=|R\rangle\langle R| - |L\rangle\langle L|$) on photon
$B$ is required. With these conditional operations, the final state
of photon $B$ is obtained as
$|\varphi_B\rangle_f=(\alpha_1|R\rangle+\alpha_2|L\rangle)_B(\delta_1|b_1\rangle+\delta_2|b_2\rangle)$.
This is just the result of the QSJM. The QSJM for other conditions
can be implemented in the same way assisted by the quantum circuit
shown in Fig.~\ref{figure3.9}b, such as combining the spatial-mode
state of photon $A$ and the spatial-mode state of photon $B$ into an
output single photon state.

\begin{center}                        
\begin{figure}[!h]
\includegraphics*[width=3in]{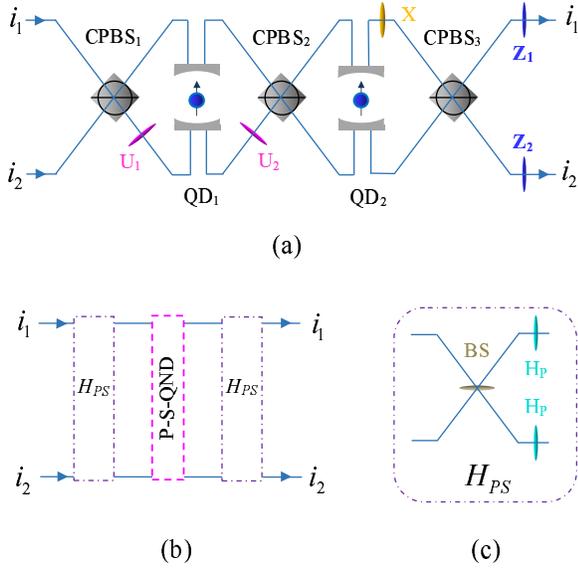}
\caption{(a) Schematic diagram of the polarization-spatial
phase-check QND (P-S-QND)  \cite{HEPP1}. Z$_i$ ($i=1,2$) represents
a wave plate which is used to perform a polarization phase-flip
operation $\sigma^P_z=-|R\rangle\langle R|+|L\rangle\langle L|$ on a
photon. (b) Schematic diagram of the polarization-spatial
parity-check QND. (c) Schematic diagram of $H_{PS}$ which is used to
perform the Hadamard operations on both the polarization and
spatial-mode DOFs of a photon. }\label{figure3.10}
\end{figure}
\end{center}

\emph{Polarization-spatial phase-check QND} --- The P-S-QND is used
to distinguish the Bell state with a relative phase $0$ from the one
with a relative phase $\pi$ in both the polarization and
spatial-mode DOFs. It is constructed with the hybrid CNOT gate
(introduced in Sec.~\ref{sec4.1}) based on the nonlinearity of
double-sided QD-cavity systems \cite{HEPP1}, as shown in
Fig.~\ref{figure3.10}a. The states of the excess electron spins
$e_1$ in QD$_1$ and $e_2$ in QD$_2$ are prepared in
$|+\rangle_{e_1}$ and $|+\rangle_{e_2}$, respectively. Here
$|\pm\rangle_e=\frac{1}{\sqrt{2}}(|\uparrow\rangle\pm|\downarrow\rangle)_e$.

After the two photons $A$ and $B$ in the hyperentangled Bell state
pass through the quantum circuit shown in Fig.~\ref{figure3.10}a
in sequence, the state of the system $ABe_1e_2$ evolves to
\begin{eqnarray}                           
|\phi^\pm\rangle_P|\phi^\pm\rangle_S|+\rangle_{e_1}|+\rangle_{e_2}&\rightarrow&|\phi^\pm\rangle_P|\phi^\pm\rangle_S|\pm\rangle_{e_1}|\mp\rangle_{e_2},\nonumber\\
|\phi^\pm\rangle_P|\psi^\pm\rangle_S|+\rangle_{e_1}|+\rangle_{e_2}&\rightarrow&|\phi^\pm\rangle_P|\psi^\pm\rangle_S|\pm\rangle_{e_1}|\mp\rangle_{e_2},\nonumber\\
|\psi^\pm\rangle_P|\phi^\pm\rangle_S|+\rangle_{e_1}|+\rangle_{e_2}&\rightarrow&|\psi^\pm\rangle_P|\phi^\pm\rangle_S|\pm\rangle_{e_1}|\mp\rangle_{e_2},\nonumber\\
|\psi^\pm\rangle_P|\psi^\pm\rangle_S|+\rangle_{e_1}|+\rangle_{e_2}&\rightarrow&|\psi^\pm\rangle_P|\psi^\pm\rangle_S|\pm\rangle_{e_1}|\mp\rangle_{e_2}.\;\;\;\;\;\;\;\;
\end{eqnarray}
Then the excess electron spins $e_1$ and $e_2$ are measured
in the orthogonal basis $\{|+\rangle_e, |-\rangle_e\}$. If the state
of the excess electron spin $e_1$ is
$|+\rangle_{e_1}$, the relative phase of the polarization
state is $0$. If the state
of the excess electron spin $e_1$ is
$|-\rangle_{e_1}$, the relative phase of the polarization
state is $\pi$. If the state
of the excess electron spin $e_2$ is
$|-\rangle_{e_2}$, the relative phase of the spatial-mode
state is $0$. If the state
of the excess electron spin $e_2$ is
$|+\rangle_{e_2}$, the relative phase of the
spatial-mode state is $\pi$. Here,
\begin{eqnarray}                           
|\phi^\pm\rangle_P&=&\frac{1}{\sqrt{2}}(|RR\rangle\pm|LL\rangle),\nonumber\\
|\psi^\pm\rangle_P&=&\frac{1}{\sqrt{2}}(|RL\rangle\pm|LR\rangle),\nonumber\\
|\phi^\pm\rangle_S&=&\frac{1}{\sqrt{2}}(|a_1b_1\rangle\pm|a_2b_2\rangle),\nonumber\\
|\psi^\pm\rangle_S&=&\frac{1}{\sqrt{2}}(|a_1b_2\rangle\pm|a_2b_1\rangle).
\end{eqnarray}

If the Hadamard operations are performed on both the spatial-mode
and polarization DOFs of photons $A$ and $B$ (as shown in
Fig.~\ref{figure3.10}b) before and after they pass through the
quantum circuit in Fig.~\ref{figure3.10}a, the result of the
polarization-spatial parity-check QND is obtained. That is, if the
state of the excess electron spin $e_1$ is $|+\rangle_{e_1}$, the
polarization state is in an even-parity mode. If the state of the
excess electron spin $e_1$ is $|-\rangle_{e_1}$, the polarization
state is in an odd-parity mode. If the state of the excess electron
spin $e_2$ is $|-\rangle_{e_2}$, the spatial-mode state is in an
even-parity mode. If the state of the excess electron spin $e_2$ is
$|+\rangle_{e_2}$, the spatial-mode state is in an odd-parity mode.

\begin{center}                        
\begin{figure}[!h]
\includegraphics*[width=2.7in]{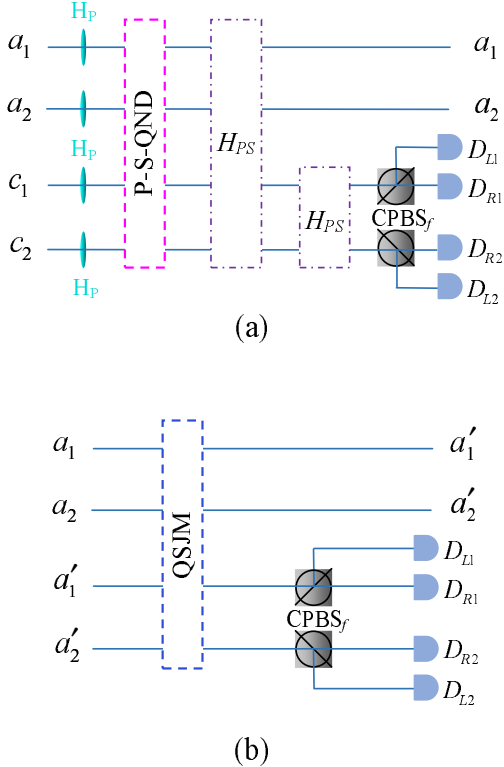}
\caption{(a) Schematic diagram of the first step of the two-step
hyper-EPP with the P-S-QND  \cite{HEPP1}. The operations performed
on the photons $B$ and $D$ are the same as the ones performed on the
photons $A$ and $C$. (b) Schematic diagram of the second step of the
two-step hyper-EPP with  QSJM. The operations performed on the
photons $B$ and $B'$ are the same as the ones performed on the
photons $A$ and $A'$.}\label{figure3.11}
\end{figure}
\end{center}

The two-step hyper-EPP for mixed hyperentangled Bell states with polarization
bit-flip errors and spatial-mode phase-flip errors is constructed with the
P-S-QND and the QSJM \cite{HEPP1}, as shown in Fig.~\ref{figure3.11}.

\emph{The first step of the hyper-EPP} ---
The setup of the first step of the hyper-EPP for mixed hyperentangled Bell states is shown in Fig.~\ref{figure3.11}a.
Here, the phase-flip errors of Bell states can be transformed to the bit-flip errors
by performing Hadamard operations on the two qubits.
Suppose that there are two identical nonlocal two-photon systems $AB$ and $CD$,
and they are in the states
\begin{eqnarray}                           
\rho_{AB}\!\!&=&\!\!\left[F_1|\phi^+\rangle_P\langle\phi^+|+(1-F_1)|\psi^+\rangle_P\langle\psi^+|\right]_{AB}\nonumber\\
&&
\!\!\otimes\left[F_2|\phi^+\rangle_S\langle\phi^+|+(1-F_2)|\phi^-\rangle_S\langle\phi^-|\right]_{AB},\nonumber\\
\rho_{CD}\!\!&=&\!\!\left[F_1|\phi^+\rangle_P\langle\phi^+|+(1-F_1)|\psi^+\rangle_P\langle\psi^+|\right]_{CD}\nonumber\\
&&\!\!
\otimes\left[F_2|\phi^+\rangle_S\langle\phi^+|+(1-F_2)|\phi^-\rangle_S\langle\phi^-|\right]_{CD}.\;\;\;\;
\end{eqnarray}
Here the subscripts $AB$ and $CD$ represent two nonlocal photon pairs.
The two photons $A$ and $C$ are obtained by Alice, and
the two photons $B$ and $D$ are obtained by Bob. $F_1$ and $F_2$ represent
the probabilities of $|\phi^+\rangle_P$ and $|\phi^+\rangle_S$ in the
mixed state $\rho$, respectively.

Initially, the four-photon system $ABCD$ is in the state
$\rho_0=\rho_{AB}\otimes\rho_{CD}$, which is a mixed  state composed
of 16 maximally hyperentangled pure states. Alice and Bob both
perform H$_P$, P-S-QND, and H$_{PS}$ operations on the polarization
and spatial-mode DOFs of their photon pairs $AC$ and $BD$. After the
measurements are performed on the excess electron spins $e_1$ and
$e_2$ in P-S-QND, the states of the four-photon systems can be
divided into four cases, which are discussed in detail as follows.

(1) The results of the P-S-QNDs show that the two photon pairs $AC$
and $BD$ are in the same polarization parity mode and the same
spatial-mode parity mode. The polarization state of the four-photon
system $ABCD$ is projected to a mixed state that consists of
$|\Psi_1\rangle_P$ and $|\Psi_2\rangle_P$ (or
$|\widetilde{\Psi}_1\rangle_P$ and $|\widetilde{\Psi}_2\rangle_P$),
and the spatial-mode state of four-photon system $ABCD$ is projected
to a mixed state that consists of $|\Psi_1\rangle_S$ and
$|\Psi_2\rangle_S$ (or $|\widetilde{\Psi}_1\rangle_S$ and
$|\widetilde{\Psi}_2\rangle_S$). Here
\begin{eqnarray}                           
&|\Psi_1\rangle_P&=\frac{1}{\sqrt{2}}(|RRRR\rangle+|LLLL\rangle),\nonumber\\
&|\Psi_2\rangle_P&=\frac{1}{\sqrt{2}}(|RLRL\rangle+|LRLR\rangle),\nonumber\\
&|\widetilde{\Psi}_1\rangle_P&=\frac{1}{\sqrt{2}}(|RRLL\rangle+|LLRR\rangle),\nonumber\\
&|\widetilde{\Psi}_2\rangle_P&=\frac{1}{\sqrt{2}}(|RLLR\rangle+|LRRL\rangle),\nonumber\\
&|\Psi_1\rangle_S&=\frac{1}{\sqrt{2}}(|a_1b_1c_1d_1\rangle+|a_2b_2c_2d_2\rangle),\nonumber\\
&|\Psi_2\rangle_S&=\frac{1}{\sqrt{2}}(|a_1b_2c_1d_2\rangle+|a_2b_1c_2d_1\rangle),\nonumber\\
&|\widetilde{\Psi}_1\rangle_S&=\frac{1}{\sqrt{2}}(|a_1b_1c_2d_2\rangle+|a_2b_2c_1d_1\rangle),\nonumber\\
&|\widetilde{\Psi}_2\rangle_S&=\frac{1}{\sqrt{2}}(|a_1b_2c_2d_1\rangle+|a_2b_1c_1d_2\rangle).
\end{eqnarray}
The state $|\widetilde{\Psi}_i\rangle_P$
($|\widetilde{\Psi}_i\rangle_S$) can be transformed to the state
$|\Psi_i\rangle_P$ ($|\Psi_i\rangle_S$) by performing the
polarization (spatial-mode) bit-flip operations on photons $C$ and
$D$. After Alice and Bob perform the Hadamard operations and
detections on the polarization and spatial-mode DOFs of the two
photons $C$ and $D$ and the conditional local phase-flip operations
$\sigma^S_z$ ($\sigma^P_z$) on photon $B$, the state of the
two-photon system $AB$ is projected to
\begin{eqnarray}                           
\rho'_{AB}\!\!&=&\!\!\left[F'_1|\phi^+\rangle_P^{AB}\langle\phi^+|+(1-F'_1)|\psi^+\rangle_P^{AB}\langle\psi^+|\right]\nonumber\\
\!\!&&\otimes\!\left[F'_2|\phi^+\rangle_S^{AB}\langle\phi^+|+(1-F'_2)|\psi^+\rangle_S^{AB}\langle\psi^+|\right]\!,\;\;\;\;\;\;\;\;
\end{eqnarray}
where $F'_i=\frac{F_i^2}{F_i^2+(1-F_i)^2}$, $F_i>1/2$($i=1,2$).

(2) The results of the P-S-QNDs show that the two photon pairs $AC$
and $BD$ are in the different polarization parity modes and the
different spatial-mode parity modes. The polarization state of the
four-photon system $ABCD$ is projected to a mixed state that
consists of $|\Psi_3\rangle_P$ and $|\Psi_4\rangle_P$ (or
$|\widetilde{\Psi}_3\rangle_P$ and $|\widetilde{\Psi}_4\rangle_P$),
and the spatial-mode state of the four-photon system $ABCD$ is
projected to a mixed state that consists of $|\Psi_3\rangle_S$ and
$|\Psi_4\rangle_S$ (or $|\widetilde{\Psi}_3\rangle_S$ and
$|\widetilde{\Psi}_4\rangle_S$). Here
\begin{eqnarray}                           
&|\Psi_3\rangle_P&=\frac{1}{\sqrt{2}}(|RRRL\rangle+|LLLR\rangle),\nonumber\\
&|\Psi_4\rangle_P&=\frac{1}{\sqrt{2}}(|RLRR\rangle+|LRLL\rangle),\nonumber\\
&|\widetilde{\Psi}_3\rangle_P&=\frac{1}{\sqrt{2}}(|RRLR\rangle+|LLRL\rangle),\nonumber\\
&|\widetilde{\Psi}_4\rangle_P&=\frac{1}{\sqrt{2}}(|RLLL\rangle+|LRRR\rangle),\nonumber\\
&|\Psi_3\rangle_S&=\frac{1}{\sqrt{2}}(|a_1b_1c_1d_2\rangle+|a_2b_2c_2d_1\rangle),\nonumber\\
&|\Psi_4\rangle_S&=\frac{1}{\sqrt{2}}(|a_1b_2c_1d_1\rangle+|a_2b_1c_2d_2\rangle),\nonumber\\
&|\widetilde{\Psi}_3\rangle_S&=\frac{1}{\sqrt{2}}(|a_1b_1c_2d_1\rangle+|a_2b_2c_1d_2\rangle),\nonumber\\
&|\widetilde{\Psi}_4\rangle_S&=\frac{1}{\sqrt{2}}(|a_1b_2c_2d_2\rangle+|a_2b_1c_1d_1\rangle).
\end{eqnarray}
In this case, Alice and Bob cannot distinguish which one of the photon pairs $AB$
and $CD$ has the polarization bit-flip error (or the spatial-mode
bit-flip error), so the two photon pairs have to be discarded.

(3) The results of the P-S-QNDs show that the two photon pairs $AC$
and $BD$ are in the same polarization parity mode and the different
spatial-mode parity modes. The polarization state of the four-photon
system $ABCD$ is projected to a mixed state that consists of
$|\Psi_1\rangle_P$ and $|\Psi_2\rangle_P$ (or
$|\widetilde{\Psi}_1\rangle_P$ and $|\widetilde{\Psi}_2\rangle_P$),
and the spatial-mode state of the four-photon system $ABCD$ is
projected to a mixed state that consists of $|\Psi_3\rangle_S$ and
$|\Psi_4\rangle_S$ (or $|\widetilde{\Psi}_3\rangle_S$ and
$|\widetilde{\Psi}_4\rangle_S$). After Alice and Bob perform the
Hadamard operations and detections on the polarization and
spatial-mode DOFs of the two photons $C$ and $D$ and the conditional
local phase-flip operations $\sigma^S_z$ ($\sigma^P_z$) on photon
$B$, the state of the two-photon system $AB$ is projected to
\begin{eqnarray}                           
\rho''_{AB}\!\!&=&\!\!\left[F'_1|\phi^+\rangle_P^{AB}\langle\phi^+|+(1-F'_1)|\psi^+\rangle_P^{AB}\langle\psi^+|\right]\nonumber\\
\!\!&&\otimes\left[F''_2|\phi^+\rangle_S^{AB}\langle\phi^+|+(1-F''_2)|\psi^+\rangle_S^{AB}\langle\psi^+|\right]\!,\;\;\;\;\;\;\;\;\;
\end{eqnarray}
where $F''_i=\frac{F_i(1-F_i)}{2F_i(1-F_i)}$. In this case,
Alice and Bob cannot distinguish which one of the photon pairs $AB$
and $CD$ has the spatial-mode bit-flip error, so the second step of
hyper-EPP is required to perform on the photon pair $AB$ to obtain the
high-fidelity hyperentangled state.

(4) The results of the P-S-QNDs show that the two photon pairs $AC$
and $BD$ are in the different polarization parity modes and the same
spatial-mode parity mode. The polarization state of the four-photon
system $ABCD$ is projected to a mixed state that consists of
$|\Psi_3\rangle_P$ and $|\Psi_4\rangle_P$ (or
$|\widetilde{\Psi}_3\rangle_P$ and $|\widetilde{\Psi}_4\rangle_P$),
and the spatial-mode state of the four-photon system $ABCD$ is
projected to a mixed state that consists of $|\Psi_1\rangle_S$ and
$|\Psi_2\rangle_S$ (or $|\widetilde{\Psi}_1\rangle_S$ and
$|\widetilde{\Psi}_2\rangle_S$). After Alice and Bob perform the
Hadamard operations and detections on the polarization and
spatial-mode DOFs of the two photons $C$ and $D$ and the conditional
local phase-flip operations $\sigma^S_z$ ($\sigma^P_z$) on photon
$B$, the state of the two-photon system $AB$ is projected to
\begin{eqnarray}                           
\rho'''_{AB}\!\!&=&\!\!\left[F''_1|\phi^+\rangle_P^{AB}\langle\phi^+|+(1-F''_1)|\psi^+\rangle_P^{AB}\langle\psi^+|\right]\nonumber\\
\!\!&&\otimes\left[F'_2|\phi^+\rangle_S^{AB}\langle\phi^+|+(1-F'_2)|\psi^+\rangle_S^{AB}\langle\psi^+|\right]\!.\;\;\;\;\;\;\;
\end{eqnarray}
In this case, Alice and Bob cannot distinguish which one of the photon pairs $AB$
and $CD$ has the polarization bit-flip error, so the second step of
hyper-EPP is required to perform on the photon pair $AB$ to obtain the
high-fidelity hyperentangled state.

\emph{The second step of the hyper-EPP} --- Here, four identical
nonlocal photon pairs $AB$, $CD$, $A'B'$, and $C'D'$ are required.
The photons $A$, $C$, $A'$, and $C'$ are obtained by Alice, and the
photons $B$, $D$, $B'$, and $D'$ are obtained by Bob. In the first
step, the same operations are performed on the four-photon systems
$ABCD$ and $A'B'C'D'$. If the results of the P-S-QNDs showed that
the four-photon systems $ABCD$ and $A'B'C'D'$ are projected to the
cases (3) and (4) in the first step, respectively, the QSJM is
introduced to combine the polarization state of the photon pair $AB$
and the spatial-mode state of the photon pair $A'B'$ into an output
single photon-pair state \cite{HEPP1}. So the preserving condition
of the case (1) in the first step is achieved.

If the results of the P-S-QNDs showed that the four-photon systems
$ABCD$ and $A'B'C'D'$ are projected to the cases (4) and (3) in the
first step, respectively, the QSJM is introduced to combine the
spatial-mode state of the photon pair $AB$ and the polarization
state of the photon pair $A'B'$ into an output single photon-pair
state. So the preserving condition of the case (1) in the first step
is also achieved.

\begin{center}                         
\begin{figure}[!h]
\includegraphics*[width=3in]{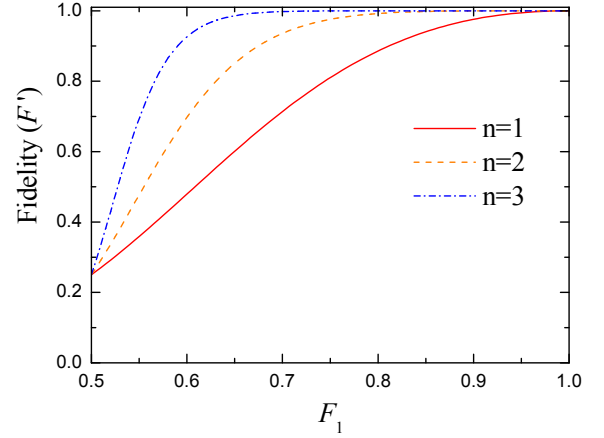}
\caption{ The fidelity of the hyperentangled Bell state obtained in
the hyper-EPP versus  the fidelity of the initial mixed
hyperentangled Bell state ($F_1$) under the iteration number ($n$)
\cite{HEPP1}. Here, the parameters of the initial  mixed
hyperentangled Bell state are $F_1=F_2$.  }\label{figure3.7}
\end{figure}
\end{center}

\emph{Fidelity and efficiency } --- After the two steps of hyper-EPP
are performed on the nonlocal photon systems, the final state of the
photon pair $AB$ is obtained as
\begin{eqnarray}                           
\rho'_{AB}\!\!&=&\!\!\left[F'_1|\phi^+\rangle^P_{AB}\langle\phi^+|+(1-F'_1)|\psi^+\rangle^P_{AB}\langle\psi^+|\right]\nonumber\\
\!\!&&
\otimes\left[F'_2|\phi^+\rangle^S_{AB}\langle\phi^+|+(1-F'_2)|\psi^+\rangle^S_{AB}\langle\psi^+|\right]\!.\;\;\;\;\;\;\;\;\;
\end{eqnarray}
The fidelity of the state
$|\phi^+\rangle^P_{AB}\otimes|\phi^+\rangle^S_{AB}$ in $\rho'_{AB}$
is increased from $F=F_1\times F_2$ to $F'=F'_1\times F'_2$. By
performing the Hadamard operations on the spatial-mode DOF of the
photons $A$ and $B$, the state $|\psi^+\rangle^S_{AB}$ can be
transformed to $|\phi^-\rangle^S_{AB}$. The fidelity of this
two-photon hyperentangled state can be greatly improved by iterative
application of the two-step hyper-EPP process, as shown in
Fig.~\ref{figure3.7}.

\begin{center}                        
\begin{figure}[!h]
\includegraphics*[width=2.7in]{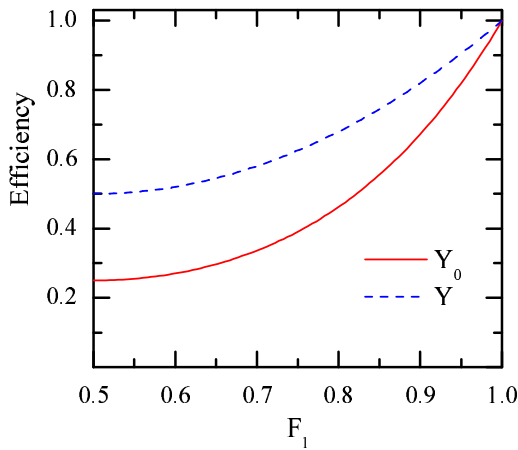}
\caption{ The efficiency of the two-step hyper-EPP for mixed
hyperentangled Bell states  \cite{HEPP1}. $Y_0$ ($Y$) is the
efficiency of the first round of the hyper-EPP process without
(with) QSJM. The parameters of the initial mixed hyperentangled Bell
state are $F_1=F_2$.}\label{figure3.12}
\end{figure}
\end{center}

The efficiency is defined as the probability to obtain a
high-fidelity entangled photon system from a pair of less-entangled
photon systems after they transmitted over a noisy channel (without
considering the photon loss). In the hyper-EPP without QSJM, only
the case (1) in  the first step is preserved, so the efficiency of
the first round in the hyper-EPP process is
\begin{eqnarray}                           
Y_0=\left[F_1^2+(1-F_1)^2\right]\times\left[F_2^2+(1-F_2)^2\right].
\end{eqnarray}
In the two-step hyper-EPP with QSJM, the high-fidelity
hyperentangled states can also be obtained from the cases (3) and
(4) in  the first step, so the efficiency of the first round in the
hyper-EPP process is
\begin{eqnarray}                           
Y=F_2^2\!+\!(1\!-\!F_2)^2.
\end{eqnarray}
Here $F_1>F_2$. Now, we can see that the efficiency of the hyper-EPP
is greatly improved by introducing the second step with  QSJM, as
shown in Fig.~\ref{figure3.12} ($F_1=F_2$).

\section{Hyperparallel photonic quantum computation }
\label{sec4.1}

Photonic quantum computation is an important branch of parallel
quantum computation. With the nonlinear interaction between a photon
and an artificial atomic system, photonic quantum gates are scalable
as the same as the universal quantum gates on solid-state quantum
systems, such as circuit quantum electrodynamics (QED) with
superconducting Josephson junctions (act as the artificial atoms)
and a superconducting resonator (acts as a cavity and quantum bus)
\cite{superconduct1,superconduct2,superconduct3,superconduct4,superconduct5,superconduct6},
diamond nitrogen-vacancy center
\cite{NVcomput1,NVcomput3,NVcomput0,NVcomput2}, quantum dots
\cite{QDcomput1,QDcomput2,QDcomput3,QDcomput4,QDcomput5,QDcomput6},
nuclear magnetic resonance
\cite{NMRcomput1,NMRcomput2,NMRcomput3,NMRcomput4}, and cavity QED
\cite{Cavity1,Cavity2}, especially with the development of the
manipulation of single photons \cite{SPM1,SPM2}. Moreover, a photon
system has multiple DOFs which can be used to encode  information in
quantum computation \cite{PhotonC}, such as polarization,
spatial-mode, frequency, orbital angular momentum, transverse,
energy-time, time bin, and so on. With the polarization DOF of
photon systems, many quantum logic gates have been constructed
either in theory or in experiment
\cite{Polar1,Polar2,Polar3,Polar4,Polar5,Polar6,Polar7,Polar8wei,Qcompadd1}.
The quantum logic gates and the quantum algorithms on a photon with
two DOFs have also been investigated in the past few years
\cite{TDOF0,TDOF1,TDOF2,TDOF3,TDOF4,TDOF5,TDOF6,TDOF7,TDOF8},
although their scalability is not good with linear or nonlinear
optical elements.

In 2013, the concept of hyperparallel photonic quantum computation,
performing universal quantum gate operations on two-photon or
multi-photon systems by encoding all the quantum states of each
photon in multiple DOFs (two or more DOFs) as information carriers,
was introduced by Ren et al. \cite{hypercnot1}. They
proposed the first scheme for the hyper-CNOT gate operating on both
the spatial-mode and the polarization DOFs of a two-photon system
simultaneously. In this scheme, both the polarization quantum state
and the spatial-mode quantum state of each photon are encoded as the
qubits for carrying information, not exploiting one DOF to implement
a CNOT gate on the other DOF of a single photon, far different from
conventional parallel quantum computation. In 2014, Ren and Deng
\cite{hypercnot} proposed another scheme for hyperparallel photonic
quantum computation assisted by the giant optical circular
birefringence induced by quantum-dot spins in double-sided optical
microcavities. It has a simpler quantum circuit. In 2015, Ren et al.
 \cite{h-hypercnot} designed two universal hyperparallel
hybrid photonic quantum logic gates with dipole-induced transparency
of a diamond NV center embedded in a photonic crystal cavity coupled
to two waveguides in the weak-coupling regime, including a hybrid
hyper-CNOT gate and a hybrid hyper-Toffoli gate on photon systems in
both the polarization and the spatial-mode  DOFs, which are equal to
two identical quantum logic gates operating simultaneously on the
two-photon systems in one DOF. Now, some important schemes for  the
hyperparallel photonic quantum computation are proposed, including
hyperparallel two-photon gates
\cite{hypercnot,hypercnot1,h-hypercnot,hypercnot4} and hyperparallel
three-photon gates (hyperparallel Toffoli gates and Fredkin gates)
\cite{h-hypercnot,hypercnot5}. With hyperparallel photonic quantum
logic gates, the resource consumed can be reduced and the photonic
dispassion noise can be depressed in quantum circuit
\cite{h-hypercnot}. Moreover, the multiple-photon hyperentangled
state can be prepared and measured with less resource and less steps
by using the hyperparallel photonic quantum logic gates, which may
speedup the quantum algorithm \cite{hypercnot,hypercnot1}.




Here, we introduce the principle of hyperparallel photonic quantum
computation by describing the process for constructing the
hyper-CNOT gate \cite{hypercnot} on both the polarization and
spatial-mode DOFs of a two-photon system, assisted by the
nonlinearity of double-sided QD-cavity systems. This gate can
achieve scalable hyperparallel quantum computation without using
auxiliary spatial modes or polarization modes.

The setup of the hyper-CNOT gate can also be described by
Fig. \ref{figure3.10}a. The states of the excess electron spins
$e_1$ in QD$_1$ and $e_2$ in QD$_2$ are prepared in
$|+\rangle_{e_1}$ and $|+\rangle_{e_2}$, respectively. The initial
states of the two photons $A$ and $B$ are
$|\varphi_A\rangle_0=(\alpha_1|R\rangle+\alpha_2|L\rangle)_A(\gamma_1|a_1\rangle
+\gamma_2|a_2\rangle)$ and
$|\varphi_B\rangle_0=(\beta_1|R\rangle+\beta_2|L\rangle)_B(\delta_1|b_1\rangle
+\delta_2|b_2\rangle)$, respectively.

First, the Hadamard operations are performed on the polarization and
spatial-mode DOFs of photon $A$, and the state of photon $A$ is
transformed to
$|\varphi'_{A}\rangle_0=(\alpha'_1|R\rangle+\alpha'_2|L\rangle)_A(\gamma'_1|a_1\rangle
+\gamma'_2|a_2\rangle)$.
Then the two spatial modes $a_1$ and $a_2$ of photon $A$ are put into
CPBS$_1$, U$_1$, QD$_1$, U$_2$, and CPBS$_2$ in
sequence as shown in Fig.~\ref{figure3.10}a, and the state of the
system $Ae_1$ is
transformed from $|\varphi_{Ae_1}\rangle_0$ to
$|\varphi_{Ae_1}\rangle_1$. Here
\begin{eqnarray}                           \label{eq.11}    
|\varphi_{Ae_1}\rangle_0&\!\!=\!\!&\frac{1}{\sqrt{2}}(|\uparrow\rangle+|\downarrow\rangle)_{e_1}\otimes\nonumber\\
&&\!\!(\alpha'_1|R\rangle+\alpha'_2|L\rangle)_A
(\gamma'_1|a_1\rangle+\gamma'_2|a_2\rangle),\nonumber\\
|\varphi_{Ae_1}\rangle_1&\!\!=\!\!&\frac{1}{\sqrt{2}}\{\gamma'_1\big[|\uparrow\rangle_{e_1}(\alpha'_1|L\rangle+\alpha'_2|R\rangle)_A
\nonumber\\
&&\!\!+|\downarrow\rangle_{e_1}(\alpha'_1|R\rangle+\alpha'_2|L\rangle)_a)\big]|a_2\rangle\nonumber\\
&&\!\!+\gamma'_2\big[|\uparrow\rangle_{e_1}(\alpha'_1|R\rangle+\alpha'_2|L\rangle)_A\nonumber\\
&&\!\!
+|\downarrow\rangle_{e_1}(\alpha'_1|L\rangle+\alpha'_2|R\rangle)_a)\big]|a_1\rangle\}.\label{eq10}
\end{eqnarray}
After the two spatial modes $a_1$ and $a_2$ pass through CPBS$_2$,
we put them into
QD$_2$, X, CPBS$_3$, Z$_1$, and Z$_2$ in sequence as shown in Fig.
\ref{figure3.10}a, and the state of the system $Ae_1e_2$  is
transformed from $|\varphi_{Ae_1e_2}\rangle_1$ to
$|\varphi_{Ae_1e_2}\rangle_2$. Here
\begin{eqnarray}                           \label{eq.12}    
|\varphi_{Ae_1e_2}\rangle_1&\!\!=\!\!&\frac{1}{\sqrt{2}}(|\uparrow\rangle+|\downarrow\rangle)_{e_2}\otimes|\varphi_{Ae_1}\rangle_1,\nonumber\\
|\varphi_{Ae_1e_2}\rangle_2&\!\!=\!\!&\frac{1}{2}\big[|\uparrow\rangle_{e_1}(\alpha'_1|R\rangle+\alpha'_2|L\rangle)_A
+|\downarrow\rangle_{e_1}(\alpha'_2|R\rangle\nonumber\\
&&\!\! +\alpha'_1|L\rangle)_A)\big]
\big[|\uparrow\rangle_{e_2}(\gamma'_2|a_1\rangle+\gamma'_1|a_2\rangle)\nonumber\\
&&\!\!
-|\downarrow\rangle_{e_2}(\gamma'_1|a_1\rangle+\gamma'_2|a_2\rangle)\big].\label{eq12}
\end{eqnarray}
Now, we have obtained the result of the four-qubit hybrid CNOT gate.

Subsequently, we perform the Hadamard operations on the electron spins
$e_1$ and $e_2$, and we put the two spatial modes $b_1$ and $b_2$ of photon $B$ into
CPBS$_1$, U$_1$, QD$_1$, U$_2$, CPBS$_2$, QD$_2$, X, CPBS$_3$,
Z$_1$, and Z$_2$ in sequence as shown in Fig.~\ref{figure3.10}a. Then the state
of the system $ABe_1e_2$ is transformed from
$|\varphi_{ABe_1e_2}\rangle_2$ to $|\varphi_{ABe_1e_2}\rangle_3$.
Here
\begin{eqnarray}                           \label{eq.13}    
|\varphi_{ABe_1e_2}\rangle_2&\!\!=\!\!&|\varphi_{Ae_1e_2}\rangle_2(\beta_1|R\rangle+\beta_2|L\rangle)_B
(\delta_1|b_1\rangle+\delta_2|b_2\rangle),\nonumber\\
|\varphi_{ABe_1e_2}\rangle_3&\!\!=\!\!&\frac{1}{2}\big[|\uparrow\rangle_{e_1}\alpha_1(|R\rangle+|L\rangle)_A(\beta_1|R\rangle+\beta_2|L\rangle)_B
\nonumber\\
&&\!\!+|\downarrow\rangle_{e_1}\alpha_2(|R\rangle-|L\rangle)_A(\beta_2|R\rangle+\beta_1|L\rangle)_B\big]\nonumber\\
&&\!\!\otimes\big[-|\uparrow\rangle_{e_2}\gamma_2(|a_1\rangle-|a_2\rangle)(\delta_2|b_1\rangle+\delta_1|b_2\rangle)
\nonumber\\
&&\!\!+|\downarrow\rangle_{e_2}\gamma_1(|a_1\rangle+|a_2\rangle)(\delta_1|b_1\rangle+\delta_2|b_2\rangle)\big].
\label{eq13}
\end{eqnarray}

Finally, the Hadamard operations are performed on the spatial-mode and
polarization DOFs of photon $A$ and the excess electron spins $e_1$
and $e_2$ again, and the state of the system $ABe_1e_2$ is changed
to
\begin{eqnarray}                           \label{eq.14}    
|\varphi_{ABe_1e_2}\rangle_4&\!\!=\!\!&\frac{1}{2}\big\{|\uparrow\rangle_{e_1}\big[\alpha_1|R\rangle_A(\beta_1|R\rangle+\beta_2|L\rangle)_B
\nonumber\\
&&\!\!+\alpha_2|L\rangle_A(\beta_2|R\rangle+\beta_1|L\rangle)_B\big]
\nonumber\\
&&\!\!
+|\downarrow\rangle_{e_1}\big[\alpha_1|R\rangle_A(\beta_1|R\rangle+\beta_2|L\rangle)_B
\nonumber\\
&&\!\!-\alpha_2|L\rangle_A(\beta_2|R\rangle+\beta_1|L\rangle)_B\big]\big\}\nonumber\\
&&\!\!
\otimes\big\{|\uparrow\rangle_{e_2}\big[\gamma_1|a_1\rangle(\delta_1|b_1\rangle+\delta_2|b_2\rangle)
\nonumber\\
&&\!\!-\gamma_2|a_2\rangle(\delta_2|b_1\rangle+\delta_1|b_2\rangle)\big]
\nonumber\\
&&\!\!-|\downarrow\rangle_{e_2}\big[\gamma_1|a_1\rangle(\delta_1|b_1\rangle+\delta_2|b_2\rangle)
\nonumber\\
&&\!\!+\gamma_2|a_2\rangle(\delta_2|b_1\rangle+\delta_1|b_2\rangle)\big]\big\}.\label{eq15}
\end{eqnarray}
Then the two excess electron spins $e_1$ and $e_2$ are measured in the
orthogonal basis $\{|\!\!\uparrow\rangle_e,|\!\!\downarrow\rangle_e\}$.
If the state of electron spin $e_1$ is
$|\downarrow\rangle_{e_1}$, an additional sign change
$|L\rangle_A\rightarrow-|L\rangle_A$ is
performed on photon $A$. If the state of electron spin $e_2$ is
$|\uparrow\rangle_{e_2}$, an additional sign change
$|a_2\rangle\rightarrow-|a_2\rangle$ is performed on photon $A$.
Now, we can obtain the result of the spatial-polarization hyper-CNOT gate,
\begin{eqnarray}                           \label{eq.15}    
|\varphi_{AB}\rangle\!\!&=&\!\!
\big[\alpha_1|R\rangle_A(\beta_1|R\rangle+\beta_2|L\rangle)_B
+\alpha_2|L\rangle_A(\beta_2|R\rangle
\nonumber\\
\!\!&&+\beta_1|L\rangle)_B \big]\otimes
 \big[\gamma_1|a_1\rangle(\delta_1|b_1\rangle+\delta_2|b_2\rangle)\nonumber\\
\!\!&& +\gamma_2|a_2\rangle(\delta_2|b_1\rangle+\delta_1|b_2\rangle)
\big].\label{eq16}
\end{eqnarray}
Here the polarization and spatial-mode DOFs of photon $A$ are used as control
qubits and the polarization and spatial-mode DOFs of photon $B$ are used
as target qubits.

\section{Discussion and Summary}

In this review, we have introduced the preparation of
hyperentanglement and its application in QIP. Hyperentanglement is
defined as the entanglement in multiple DOFs of photon system, and
it can be prepared with the combination of the techniques used for
creating the entanglement in a single DOF
\cite{preparation2,preparation3,preparation4,preparation5,preparation6,preparation7,preparation9,preparation10}.
In quantum communication, hyperentanglement can be used to increase
the channel capacity largely, besides its application for assisting
the implementation of quantum communication protocols based on one
DOF. HBSA is the prerequisite for quantum communication protocols
with hyperentanglement, and it is one of the important parts in
high-capacity quantum repeaters
\cite{HBSA,HBSA1,HBSA2,HBSA3,HBSA4,HBSALiuq,HBSA6,HBSALIXHOE,HBSA7,HBSAWW}.
We have reviewed the high-capacity long-distance quantum
communication protocols based on polarization-spatial
hyperentanglement, including the complete HBSA scheme with the
cross-Kerr nonlinearity, the quantum teleportation of a quantum
state in both the polarization and the spatial-mode DOFs with
polarization-spatial hyperentanglement, and the hyperentanglement
swapping of polarization-spatial hyperentangled Bell states. They
are useful tools in high-capacity quantum communication protocols
and high-capacity quantum repeaters.


In  practical applications, the maximally entangled photon systems
are produced locally, which leads to the decoherence of the photon
systems  by the environment noise in their distribution and storage
processes in QIP. To depress this decoherence, entanglement
concentration
\cite{HECP,HEPPECP,HECP1,HECP2,HECP4,HECP3,HECPLixhOE,HECPadd2} and
entanglement purification
\cite{HEPPECP,HEPP1,HEPP2wang,HEPPadd,HEPPaddWW}, two passive ways
for nonlocal quantum systems to overcome the adverse influence from
noise, were introduced for long-distance quantum communication
assisted by quantum repeaters
\cite{BB84,QKD2,QKD3,QSS1,QSDC1,QSDC2,QSDCDL04,DL04Exp,twostepExp,Qcommunadd1,Qcommunadd2}.
Hyper-ECPs are used to obtain the maximally hyperentangled states
from nonlocal partially hyperentangled pure states. The hyper-ECPs
can be implemented with two methods: the parameter-splitting method
\cite{HECP} and the Schmidt projection method  \cite{HECP,HEPPECP},
which are useful for improving the entanglement of the partially
hyperentangled states with their parameters known and unknown to the
remote users, respectively. In a practical quantum communication,
the information about the parameters of a nonlocal partially
hyperentangled pure state can be obtained by measuring an enough
number of sample photon pairs, and the parameter-splitting method is
far more efficient than the Schmidt projection method when they are
used to obtain maximally entangled states in the case with a large
number of quantum data needed to be exchanged between the two
parties. While, if there are a small quantity of quantum data needed
to be exchanged between the two parties, the hyper-ECPs with the
Schmidt projection method may be more practical as they do not
require the two parties to measure the samples for obtaining the
accurate information about the parameters of the partially
hyperentangled pure state. In contrast with hyper-ECPs, hyper-EPPs
are more general but they work with a relatively low efficiency. In
principle,  hyper-EPPs are used to obtain the high-fidelity
hyperentangled states from nonlocal mixed hyperentangled states with
less entanglement, not a pure state. As an example, we introduced
the two-step hyper-EPP for mixed hyperentangled Bell states with
both the polarization bit-flip errors and the spatial-mode
phase-flip errors, resorting to the nonlinearity of double-sided
QD-cavity system. The P-S-QND and QSJM are two basic quantum gate
operations for the two steps, respectively. By introducing the QSJM,
the efficiency of the hyper-EPP can be improved by preserving the
states that are discarded in the hyper-EPP without QSJM. The
fidelity of the two-photon state in hyper-EPP can be improved by
iterative application of the hyper-EPP process. This hyper-EPP can
be generalized to purify the mixed hyperentangled GHZ states with
channel errors in both the two DOFs of photon systems.

As another important application of hyperentanglement, we introduced
a photonic hyper-CNOT gate for the hyperparallel quantum
computation. The hyperparallel quantum logic gate can be used to
perform multiple quantum logic gate operations on a photon system
compared with those on multiple photon systems
\cite{hypercnot,hypercnot1,h-hypercnot,hypercnot4,hypercnot5}.
Recently, the hyperparallel photonic Toffoli gate for a three-photon
system has also been proposed \cite{h-hypercnot}, which can perform
double Toffoli gate operations on a three-photon system. With the
hyperparallel quantum logic gate, the hyperentangled states of
multiple-photon system can be prepared and analyzed with less
resource and steps, which may reduce the resource and steps required
for quantum algorithms. Moreover, the hyperparallel quantum logic
gates can be used to implement the universal hyperparallel quantum
computation tasks on multiple DOFs of photon systems, together with
the single photon manipulation.

Now, more and more attention is focused on hyperentanglement and its
applications in QIP. Maybe it is also the important resource for
other tasks in quantum physics and quantum techniques. Also, it may
be interesting to investigate hyperentanglement in relativistic
systems \cite{relativehyperentanglement}. The continuous-variable
hyperentanglement of polarization and orbital angular momentum has
been generated experimentally with a type-II optical parametric
oscillator \cite{CVhyper}.

\section*{Acknowledgment}

This work was supported by the National Natural Science Foundation
of China (11474026, 11574038,  11547106, 11604226, and 11674033).

\section*{Conflict of interests}
The authors declare that they have no conflict of interests.

\end{document}